\documentclass[iop,twocolappendix,numberedappendix]{emulateapj}

\usepackage{amsmath}
\usepackage{txfonts}
\usepackage{graphicx}
\usepackage{color}
\usepackage{hyperref}
\usepackage[utf8]{inputenc}

\newcommand{\closs}{\mathcal{C}^{\textrm{loss}}}
\newcommand{\cgain}{\mathcal{C}^{\textrm{gain}}}
\newcommand{\rmax}{r_{\textrm{max}}}
\newcommand{\rmin}{r_{\textrm{min}}}
\definecolor{RED}{rgb}{1,0,0}
\begin{document}

\title{On the Occurrence of Crossings Between the Angular Distributions of Electron Neutrinos and Antineutrinos in the Supernova Core}
\shorttitle{Crossings Between the Angular Distributions of Neutrinos in the Supernova Core}

\shortauthors{Shalgar and Tamborra}

\author{Shashank Shalgar\altaffilmark{1} and Irene Tamborra\altaffilmark{1}}

\affil{\altaffilmark{1} Niels Bohr International Academy and DARK, Niels Bohr Institute, University of Copenhagen, Blegdamsvej 17, 2100, Copenhagen, Denmark}

%\email{shashank.shalgar@nbi.ku.dk}
%\email{tamborra@nbi.ku.dk}

\begin{abstract}
Neutrino fast pairwise conversions have been postulated to occur in the dense core of a core-collapse supernova (SN), possibly having dramatic consequences  on the SN mechanism and the observable neutrino signal. One crucial condition favoring pairwise conversions is the presence of crossings between the electron neutrino and antineutrino angular distributions (i.e., electron neutrino lepton number crossings, ELN crossings). A stationary and spherically symmetric SN toy-model  is constructed to reproduce the development of the neutrino angular distributions in the dense SN core in the absence of perturbations induced by hydrodynamical instabilities. By  iteratively solving the neutrino Boltzmann equations including the collisional term,  our model predicts  that ELN crossings can develop only in the proximity of the decoupling region and for a  sharp radial evolution of the baryon density, when the electron neutrino and antineutrino number densities are comparable. Such conditions are likely to occur only in the late SN stages. Interestingly,  flavor instabilities induced by spatial or temporal perturbations are unlikely to generate ELN crossings dynamically within our simplified setup. 
\end{abstract}


\section{Introduction}
Core collapse supernovae (SNe) are among the densest sources in neutrinos~\citep{Mirizzi:2015eza,Janka:2012wk}. According to our current understanding, neutrinos are emitted as matter accretes on the proto-neutron star; they transport energy to finally revive the stalled shock-wave and power the  SN explosion~\citep{Janka:2017vcp}. 
However, a detailed picture of the role of neutrinos remains unclear.

Understanding the impact of neutrino flavor conversions on the SN inner working is one of the crucial unsolved problems. In fact, neutrino flavor conversions have been traditionally neglected in SN neutrino transport, since they were expected to occur at radii larger than the shock radius and assumed to have a negligible impact on the shock revival;  see e.g.~\cite{Dasgupta:2011jf}  where effects coming from $\nu$--$\nu$ interactions were included in hydrodynamical simulations, albeit not in a self-consistent setup. 

Recent work suggests  that  the large density of neutrinos in the proximity of the neutrino decoupling region may favor the development of pairwise conversions~\citep{Sawyer:2015dsa,Sawyer:2008zs,Izaguirre:2016gsx}. Contrary to our intuition, neutrino pairwise conversions could occur even in the absence of a hierarchy among the neutrino mass eigenstates and are independent on the neutrino energy; however, fast pairwise conversions crucially depend on the exact shape of the neutrino angular distributions and on the local neutrino density. 
The scale ruling such phenomenon is determined by $(G_F |n_{\nu_e}-n_{\bar{\nu}_e}|)^{-1} \simeq \mathcal{O}(10)$~cm, with $G_F$ being the Fermi constant and $n_{\nu_e(\bar{\nu}_e)}$ the local (anti)neutrino number density. As a direct consequence of the ``fast'' interaction rate, neutrino pairwise conversions could possibly lead to flavor decoherence~\citep{Abbar:2018beu,Dasgupta:2016dbv,Capozzi:2018clo},  affecting the neutrino heating  within the gain radius as well as the SN nucleosynthesis, and the neutrino signal observable at Earth.

 The criteria leading to the development of fast pairwise conversions were investigated in Refs.~\citep{Izaguirre:2016gsx,Capozzi:2018rzl,Capozzi:2017gqd}. 
A non-negligible flux of neutrinos streaming backwards towards the proto-neutron star core as well as  the presence of a crossing between the angular distributions of electron neutrinos and antineutrinos have been pinpointed as crucial ingredients to trigger fast  instabilities in the neutrino flavor space. A crossing between the angular distributions of $\nu_e$ and $\bar{\nu}_e$, i.e. a change of sign of $(n_{\nu_e}-n_{\bar{\nu}_e})$ within a specific  angular range, will be throughout dubbed electron neutrino lepton number crossing (ELN crossing).

 While ELN crossings are bound to occur in the case of compact binary mergers because of the particular merger geometry~\citep{Wu:2017qpc,Wu:2017drk}; the situation seems to be more complex  in the SN context.  In fact, self-consistent spherically symmetric SN simulations in 1D did not find any evidence for ELN crossings over a range of progenitor masses~\citep{Tamborra:2017ubu}. On the other hand, multi-D hydrodynamical simulations have reported mixed results~\citep{Abbar:2018shq,Azari:2019jvr} depending on the different degree of approximation adopted in the implementation of the neutrino transport. In particular, the occurrence of LESA, the lepton-emission self-sustained asymmetry~\citep{Tamborra:2014aua,Tamborra:2014hga} has been recently confirmed by a larger set of independent simulations~\citep{Janka:2016fox,Walk:2019ier,Vartanyan:2018iah,OConnor:2018tuw,Glas:2018vcs}. LESA has been also invoked as a possible ingredient favoring the development of the ELN crossings, see e.g.~\cite{Izaguirre:2016gsx,Dasgupta:2016dbv}. 

The recent developments described above hint towards the need of a full implementation of the neutrino quantum kinetic equations, including collisions and flavor conversions~\citep{Volpe:2013jgr,Stirner:2018ojk,Sigl:1992fn,Vlasenko:2013fja,Blaschke:2016xxt,Cirigliano:2017hmk}. However, given the major numerical complications involved in the problem,  this has not been tackled numerically in a self-consistent manner yet; see \cite{Richers:2019grc} for a recent simplified attempt in this direction which does not include fast pairwise conversions.

Hydrodynamical simulations of the core collapse do not include the physics of neutrino flavor conversions. Despite that, the simulations tracking the evolution of the neutrino angular distributions, see e.g.~Refs.~\citep{Tamborra:2017ubu,Sumiyoshi:2014qua,Nagakura:2017mnp}, are extremely expensive in terms of computational resources. As a consequence, it is challenging to explore the conditions leading to the development of ELN crossings and to track to the behavior of the neutrino angular distributions in the late stages of the core collapse.

In this work, by assuming spherical symmetry and that a stationary solution is reached, we develop a simplified model to mimic the SN core microphysics and   iteratively solve the neutrino Boltzmann equations with the inclusion of the collisional term. Our goal is  to estimate the neutrino angular distributions of $\nu_e$ and $\bar{\nu}_e$ in the neutrino trapping region and follow their evolution until neutrinos are fully decoupled in the different SN phases.  By neglecting complications coming from the SN hydrodynamics, we aim at identifying the microphysical ingredients leading to the development of ELN crossings and discuss their implications for the physics of flavor conversions. 

The outline of the paper is as follows. In Sec.~\ref{sec:collisions}, we present our stationary and spherically symmetric SN model and the iterative method employed to solve the neutrino Boltzmann equations by taking into account the collision term. In Sec.~\ref{sec:crossings}, we investigate the occurrence of ELN crossings by introducing a simple evaluation criterion; we first test the latter for a simplified scenario and then discuss its implications through a more realistic framework based on inputs from a 1D  hydrodynamical SN  model. Section~\ref{sec:oscillations} focuses on exploring whether ELN crossings can be generated dynamically by the occurrence of neutrino flavor conversions because of spatial or temporal perturbations. A discussion on our findings and conclusions are reported in Sec.~\ref{sec:conclusions}.

\section{Neutrino Equations of Motion in the dense proto-neutron star}
\label{sec:collisions}
In this Section, we first introduce the neutrino kinetic equations and then describe the method employed to solve them in spherical symmetry and derive the neutrino angular distributions.

\subsection{Neutrino Kinetic Equations} 
In the interior of a SN, the neutrino number density as well as the baryon density are very large and neither  neutrino-neutrino self-interactions (coherent forward scattering) nor  neutrino-matter scatterings (incoherent scattering) can be ignored. 
The coherent and incoherent parts of the neutrino evolution can have a significant feedback on each other. In fact, the neutrino-matter cross-section is flavor dependent, while the non-linear neutrino flavor evolution depends on the spatial distribution of neutrinos, which is determined by the neutrino matter scattering cross-section.  However, the interplay between the non-linearity arising from the neutrino self-interaction and scatterings with the nuclear matter remains enigmatic. 

Throughout this paper, for the sake of simplicity, we assume that ($\nu_e, \nu_x$) describes the neutrino flavor basis with $\nu_x$ being a mixture between $\nu_\mu$ and $\nu_\tau$ and similarly for antineutrinos. 
The equation of motion describing the neutrino flavor evolution, including coherent and incoherent components, can be conveniently expressed in terms of a $2\times 2$ density matrix, $\rho$ (and $\bar{\rho}$ for antineutrinos).  

The  neutrino kinetic equation, for each neutrino momentum mode $\vec{p}$, is given by
\begin{eqnarray}
\label{geneom1}
& & i\left(\frac{\partial \rho(\vec{x},\vec{p},t)}{\partial t} + \vec{v}\cdot \nabla \rho(\vec{x},\vec{p},t)
+ \vec{F} \cdot \nabla_{p} \rho(\vec{x},\vec{p},t) \right)\nonumber\\
&=& [H,\rho(\vec{x},\vec{p},t)] + i \mathcal{C}[\rho(\vec{x},\vec{p},t),\bar{\rho}(\vec{x},\vec{p},t)]\ ,
\end{eqnarray}
where $\vec{x}$ and $t$ describe the spatial and temporal coordinates, and $\vec{v}$ is the neutrino velocity. 
The Hamiltonian $H$ on the right-hand-side of the Eq.~\ref{geneom1}  contains three components, a term embedding the neutrino mixing in vacuum, a term describing the interactions of neutrinos with the SN matter background, and a term describing the neutrino-neutrino interactions, see e.g.~\cite{Mirizzi:2015eza}. Since we are interested in exploring the development of the ELN crossings, we will  neglect neutrino flavor conversions hereafter, unless otherwise stated.  
The term $\mathcal{C}[\rho,\bar{\rho}]$ describes the scattering of neutrinos with the medium. 

The left-hand-side of the Eq.~\ref{geneom1} contains the convective term, which can be reduced to
[$\cos\theta {\partial \rho(\vec{x},\vec{p},t)}/{\partial r}$]
under the assumption of spherical symmetry and stationarity.  It should be noted that,  due to the non-linearity of the neutrino-neutrino term, the convective term may lead to spontaneous breaking of spherical symmetry~\citep{Raffelt:2013rqa,Duan:2014gfa}. For the sake of simplicity, in the following, we will assume that spherical symmetry is preserved and  ignore the convective term. The term $\vec{F} \cdot \nabla_{p} \rho$ tracks the effect of external forces on the neutrino field; we will  assume that the latter has a negligible contribution. Hereafter we will also use $\hbar = c = 1$.

The  time-dependent density matrix, $\int \rho(\vec{x},\vec{p},t) d\Omega$ (with $d\Omega = d\cos\theta d\phi$ being the differential solid angle element), is normalized such that the terms on the diagonal give the total number density of neutrinos for fixed $(\vec{r},t)$. Notably, 
deep in the SN core the electron (anti-)neutrinos are highly degenerate and  are in thermal equilibrium with the SN medium. Therefore, the (anti-)neutrino number densities are well described by a Fermi-Dirac distribution:
\begin{equation}
\label{eq:F-D}
\frac{dn_{\nu}}{dE} = \frac{1}{(2 \pi)^2} \frac{E^2}{e^{(E-\mu_{\nu})/T} + 1}\ ,
\end{equation}
where the $\mu_{\nu}$ is the neutrino chemical potential, such that $\mu_{\nu_e}  = - \mu_{\bar{\nu}_e}$ and $\mu_{\nu_x}  =  \mu_{\bar{\nu}_x} = 0$, and $T$ is the temperature of the SN matter. The off-diagonal terms of the density matrix, instead,   being related to the off-diagonal terms of the evolution operator,  provide a measure of quantum coherence that determines the flavor transition probabilities. 

Solving Eq.~\ref{geneom1} may not appear to be a daunting task at first glance. However, it should be noted that the collision term $\mathcal{C}[\rho, \bar{\rho}]$ (and $\bar{\mathcal{C}}[\rho, \bar\rho]$) entails a  six dimensional integral for each $(\vec{x}, \vec{p}, t)$. The partial differential equation cannot be solved using standard discretization techniques in a 7D  space due to the limitations of available computational power. Imposing symmetries on the system may be misleading due to spontaneous breaking of symmetries in momentum and real space~\citep{Mirizzi:2015eza}. 

The exact shape of the neutrino angular distributions in the SN core depends on the interactions described by the collision term. 
 In general, the collision term, $\mathcal{C}[\rho, \bar{\rho}]$, contains two components: the loss and gain terms. The loss term accounts for the reduction of the neutrino number flux for a given direction and momentum due to scattering to another direction and momentum;  the gain term, instead, accounts for scattering of neutrinos into a given direction and momentum state. The exact form of the collision term adopted in this paper will be described in the next subsection.  As we will discuss later, because of the neutrino-matter interactions described by the collision term, the radial profile of the baryon density strongly affects the development of ELN crossings.

\subsection{Stationary and Spherically Symmetric Supernova  Model}
\label{sec:SNmodel}

In this Section, we introduce the stationary and spherically symmetric model that we develop to explore under which conditions the growth of ELN crossings is favored. Our model is based on the following assumptions:
\begin{enumerate}
\item{The number of neutrinos is locally conserved.  Absorption and emission are thus treated as effective isotropic elastic scatters.}
\item{Energy-averaged, flavor-dependent neutrino distributions are adopted.}
\item{Only two neutrino flavor eigenstates are considered.}
\item {Any dependence on  $t$ and  on the azimuthal angle $\phi$ is discarded.}
\end{enumerate}
\begin{figure}
\includegraphics[width=0.49\textwidth]{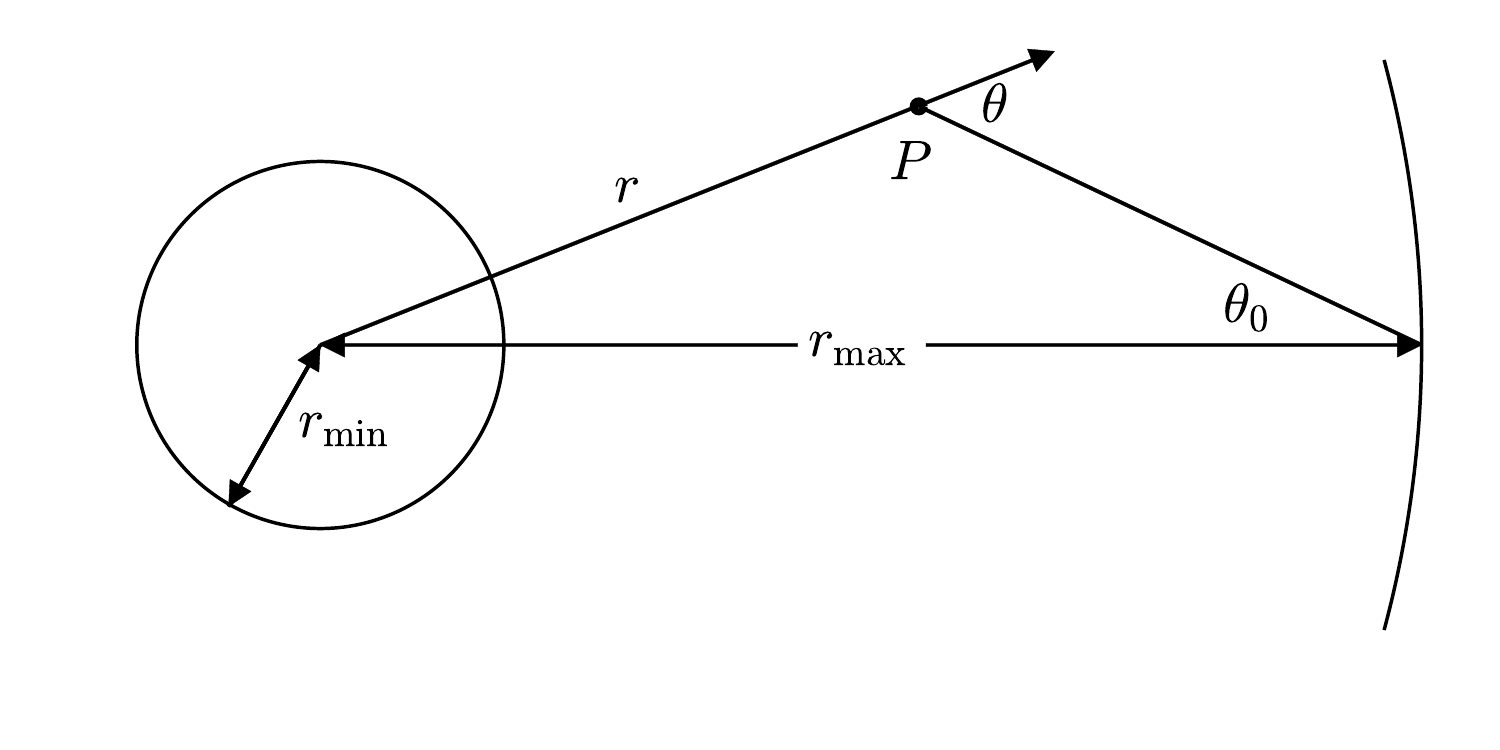}
\caption{Schematic representation of the stationary and spherically symmetric SN model. The boundary conditions are set at the innermost radius $\rmin$ and at the outermost radius $\rmax$. Each point $P$  is characterized by a global set of coordinates ($r$,  $\theta_0$) with $\theta_0$ defined with respect to the normal to the surface. For each point $P$, we also introduce a local system of coordinates through the angle    $\theta$. The angles $\theta$ and $\theta_0$ are related through Eq.~\ref{eq:angles}.}
\label{cartoon}
\end{figure}

Our spherically symmetric model is  sketched in Fig.~\ref{cartoon}. We assume that neutrinos are radiated by an inner surface of radius $\rmin$ and propagate until they reach an outermost surface of radius $\rmax$.  Each point $P$ in the SN sphere is characterized by the radius $r$ and the angle $\theta_0$, where  $\theta_{0}$ has been defined with respect to the outermost surface. For each $P$, we further introduce a set of coordinates $\theta$ to characterize the local angular distribution of neutrinos.  
The angle in the global coordinate system $\theta_{0}$  is related to the local angle $\theta$ by the following relation:
\begin{eqnarray} 
\label{eq:angles}
\cos\theta = \sqrt{1-\left(\frac{\rmax}{r}\right)^{2}\left(1-\cos^{2}\theta_{0}\right)}\ .
\end{eqnarray}
For fixed $P$, if $\mathrm{Im}(\cos\theta) \neq 0$,  then  the correspondent neutrino trajectory is discarded (i.e., the evolution equation  is not evolved along  that particular trajectory for radii such that $\mathrm{Im}(\cos\theta) \neq 0$).

From Eq.~\ref{geneom1} and  Fig.~\ref{cartoon}, one can immediately realize that we are not dealing with an initial value problem and  need  to define boundary conditions instead of initial conditions. To obtain a stationary solution in the presence of collisions, we will adopt an iterative method as described in the following.  

Given that the energy-dependent quantities entering the neutrino equation of motion are averaged over the neutrino energy distribution, we will effectively solve Boltzmann equations depending on the neutrino scattering angle, but not on their momentum.
In the first step of our iteration method, we  ignore the backward flux of neutrinos (i.e., assume a null neutrino flux for $\cos\theta<1$) and evolve the following equation of motion 
\begin{eqnarray}
\cos\theta_i \frac{d\rho^{\uparrow}_{i}(r)}{dr} &=& \sum_{j}\left(-\closs \rho^{\uparrow}_{i}(r) + \cgain \rho^{\uparrow}_{j}(r)\right) \frac{\Delta \cos\theta_{j}}{2}\ ,
\label{iter1}
\end{eqnarray}\\
where the subscripts ($i$, $j$) denote the indexes of the angular bins. $\Delta\cos\theta_{j}$ is the width of the $j^{\textrm{th}}$ angular bin, which depends on the radius and is calculated at each radial step, as we will see in the following. Note that since we are considering only effective scatterers and no flavor evolution, the  collision integral  simplifies to $\mathcal{C}[\rho,\bar{\rho}] \sim C \times \rho$ for the  loss and the gain term. The $\cos\theta_i$-term on the left hand side of Eq.~\ref{iter1}  takes into account the dependence of the path length on the zenith angle.

In order to locally conserve the neutrino  number, we set  the loss coefficient, $\closs$, equal to the gain coefficient, $\cgain$. The loss and gain coefficients are proportional to the product of the number of  effective scatterers and the cross-section for each interaction channel averaged over the neutrino energy distribution, i.e.~$\closs, \cgain = F_B n_{\mathrm{targets}} \langle \sigma \rangle$, with  $\langle \sigma \rangle$ the cross section averaged over the neutrino energy distribution:
\begin{eqnarray}
\langle \sigma \rangle = \sum_{m} \frac{\int_{E_{\mathrm{min}}}^{E_{\mathrm{max}}}\sigma_{m}(E) \frac{dn_{\nu}}{dE} dE}{\int_{E_{\mathrm{min}}}^{E_{\mathrm{max}}} \frac{dn_{\nu}}{dE} dE},
\end{eqnarray}
where the sum over $m$ is a sum over all the channels discussed below, and $[E_{\mathrm{min}},E_{\mathrm{max}}]=[0, 500]$~MeV.
The cross-sections entering $\closs$ and $\cgain$ have been calculated for the  following reactions:
\begin{eqnarray}
\nonumber
n + \nu(\bar{\nu}) &\leftrightarrow& n + \nu(\bar{\nu})\ ,\\ \nonumber
p + \nu(\bar{\nu}) &\leftrightarrow& p + \nu(\bar{\nu})\ ,\\ \nonumber
\nu(\bar{\nu}) + e^\pm &\leftrightarrow& \nu(\bar{\nu}) + e^\pm\ ,\\ \nonumber
n + e^+  &\leftrightarrow& p + \bar{\nu}_e\ ,\\ \nonumber
p + e^-  &\leftrightarrow& n + \nu_e\ .\\ \nonumber
\end{eqnarray}
Each cross-section has been  implemented as prescribed in \cite{bowers:1982}   with the low density approximation adopted for the neutrino-electron cross section. We also  employed flavor-dependent neutrino distributions defined along the lines of Eq.~\ref{eq:F-D}, and the respective mass-dependent Fermi-Dirac distributions have been adopted for the nucleons. In particular, the local density of nucleons has been defined through the baryon density $\rho_B$ and the electron fraction $Y_e$ [$n_n \simeq \rho_B (1-Y_e)$ for neutrons and $n_p \simeq \rho_B Y_e$ for protons].  
Note that, although neutrinos and antineutrinos undergo the same kind of neutral current interactions, a small difference appears in their cross sections. As we will discuss afterwords, this will also affect the eventual development of ELN crossings.  For each reaction, the Pauli blocking factor $F_B$ of scattering targets has been computed following Refs.~\citep{Raffelt:1996wa,Bruenn:1985en} and included in the collision term. Notably, in the deepest regions of the proto-neutron star, $\closs$ and $\cgain$ are suppressed by  Pauli blocking effects, however as it will be shown in the following the ELN crossings occur in the proximity of the decoupling region and therefore  Pauli blocking effects do not affect our results.

Equation~\ref{iter1} is solved from $\rmin$, arbitrarily set to 5 km, up to $\rmax$, fixed at 60 km. Note that the choice of $\rmin$ guarantees that the collisional rate is large enough to re-distribute the neutrinos in $\theta$ homogeneously, while $\rmax$ has been fixed outside of the neutrino trapping region and is  large enough to not affect the neutrino angular distributions at decoupling. During each iteration, the resulting flux of $(e,x)$ neutrinos and antineutrinos  is stored at  100 radial points per km for each angular bin.

The the loss term in Eq.~\ref{iter1}  depends  on the number of neutrinos in the $i^{\textrm{th}}$ bin and on the phase space.  However, we perform the integration  numerically to ensure that the numerical error arising from the discretization of the gain term is canceled by that of the loss term.

We  bin the neutrino trajectories  with respect to the global angle defined at $\rmax$ denoted by $\theta_{0}$. In  the numerical runs, the angular grid has been chosen to be uniform in $\cos^{2}\theta_{0}$. 
For all the results presented in this paper 200 angular bins in the global coordinate system have been used. 
We tested the convergence of our results by increasing the number of bins and made sure that the final angular distributions are stable with respect to the resolution adopted in the numerical runs within variations of less than a percent. 
 Since we use an adaptive embedded Prince-Dormand(8,9) method~\citep{contributors-gsl-gnu-2010} to solve the differential equations, the components of the density matrix between the 100 radial grid points for which the data have been stored are obtained through  linear interpolation.

After the completion of this first step of our iteration method, we evolve the equations of motion for the backward flux from $\rmax$ to $\rmin$
\begin{eqnarray}
\cos\theta_i\frac{d\rho^{\downarrow}_{i}(r)}{dr} &=& \sum_{j}\left(-\closs \rho^{\downarrow}_{i}(r) + \cgain \rho^{\downarrow}_{j}(r)\right) \frac{\Delta \cos \theta_{j}}{2}
\nonumber\\ 
&+& \sum_{j^{\prime}}\left(-\closs \rho^{\downarrow}_{i}(r) + \cgain \rho^{\textrm{pr}\uparrow}_{j^{\prime}}(r)\right) \frac{\Delta\cos\theta_{j^{\prime}}}{2}\ .
\label{iter1back}
\end{eqnarray}
Here, the superscript `pr' is used to denote that the values for the components of the density matrix are interpolated using the solutions of Eq.~\ref{iter1}. The width of the angular bins, indexed by $j$ in the forward direction and  $j^{\prime}$ in the backward direction, is the same for forward and backward going neutrinos ($\Delta\cos\theta_{j} = \Delta\cos\theta_{j^{\prime}}$). For the sake of clarity, we have divided the right hand side of Eq.~\ref{iter1back}  in two parts, each accounting for  the phase space in the forward  and backward direction respectively.
In this step, the gain term in Eq.~\ref{iter1back} receives a contribution from the forward flux stored during the previous iteration.

Upon completion of the first iteration (forward and backward), we obtain the  initial conditions of the equations of motion for the next iteration round. Since $\rmin$ is chosen to be in a region of extremely large matter density, the angular distribution is essentially uniform and we can set the forward flux at $\rmin$ equal to the backward flux for all angular bins. 
After each backward iteration, we  rescale the normalization of the backward flux by an amount that is proportional to the flux at $\rmin$ to compensate for the loss of neutrinos by diffusion at $r_{\textrm{max}}$, and achieve a steady state solution. 
The relative normalization between $\nu_{e}$ and $\bar{\nu}_{e}$ is thus determined by the dynamics of collisions in our model;  
 hence the number density at $\rmin$ is determined by the mean free path -- a smaller mean free path implies a larger number of neutrinos of a certain flavor.
This is a crucial aspect, as the occurrence of ELN crossings (or lack thereof) is determined in large part by the relative normalization of the $\nu_{e}$ and $\bar{\nu}_{e}$ angular distributions.

Using the initial conditions obtained by the former backward iteration, the equations of motion are then evolved in the forward direction, while using the interpolated values from the solution of Eq.~\ref{iter1back}:
\begin{eqnarray}
\cos\theta_i\frac{d\rho^{\uparrow}_{i}(r)}{dr} &=& \sum_{j}\left(-\closs \rho^{\uparrow}_{i}(r) + \cgain \rho^{\uparrow}_{j}(r)\right) \frac{\Delta\cos\theta_{j}}{2} \nonumber\\
&+& \sum_{j^{\prime}}\left(-\closs \rho^{\uparrow}_{i}(r) + \cgain \rho^{\textrm{pr}\downarrow}_{j^{\prime}}(r)\right) \frac{\Delta\cos\theta_{j^{\prime}}}{2}\ .
\label{iter2}
\end{eqnarray}

It should be noted that if we add Eqs.~\ref{iter1back} and \ref{iter2}, multiply by $\Delta \cos\theta_{i}$ and sum over the index $i$, the coefficient of $\closs$ and $\cgain$ is the same apart from the sign. This implies that the change in the total number of neutrinos is zero.

We repeatedly solve Eqs.~\ref{iter1back} and \ref{iter2} for neutrinos and antineutrinos using  interpolated values for the components of the density matrix that have the superscript  `pr'. 
We find that around 15 iterations are sufficient to achieve numerical convergence of the results. Notably, this procedure guarantees that  the  steady state solution is independent on the initial conditions used in the first iteration and determined by self-consistency alone.

In order to be certain that our simple SN model reproduces flavor-dependent neutrino angular distributions in agreement with the literature, we tested that our simple model gives numerical results in agreement with the analytical ones presented in~\cite{Murchikova:2017zsy} for the case of a uniform sphere emitting blackbody radiation, see Appendix for details. 

To test a more realistic scenario, we adopted the inputs of the SN models employed in Ref.~\citep{Tamborra:2017ubu} and computed the expected angular distributions. Our results are qualitatively in very good  agreement with the evolution of the neutrino angular distributions presented in \cite{Tamborra:2017ubu}, e.g.~their Figs.~4 and 6. However, given the approximate modeling of the collision term, we can reproduce well the evolution as a function of the radius of the angular distributions for each flavor as well as the relative normalization among the angular distributions of different flavors, but not their absolute normalization. Hence, in the following we will show the neutrino angular distributions  normalized to the total number of neutrinos which still serves our purposes.

\begin{figure}
\includegraphics[width=0.49\textwidth]{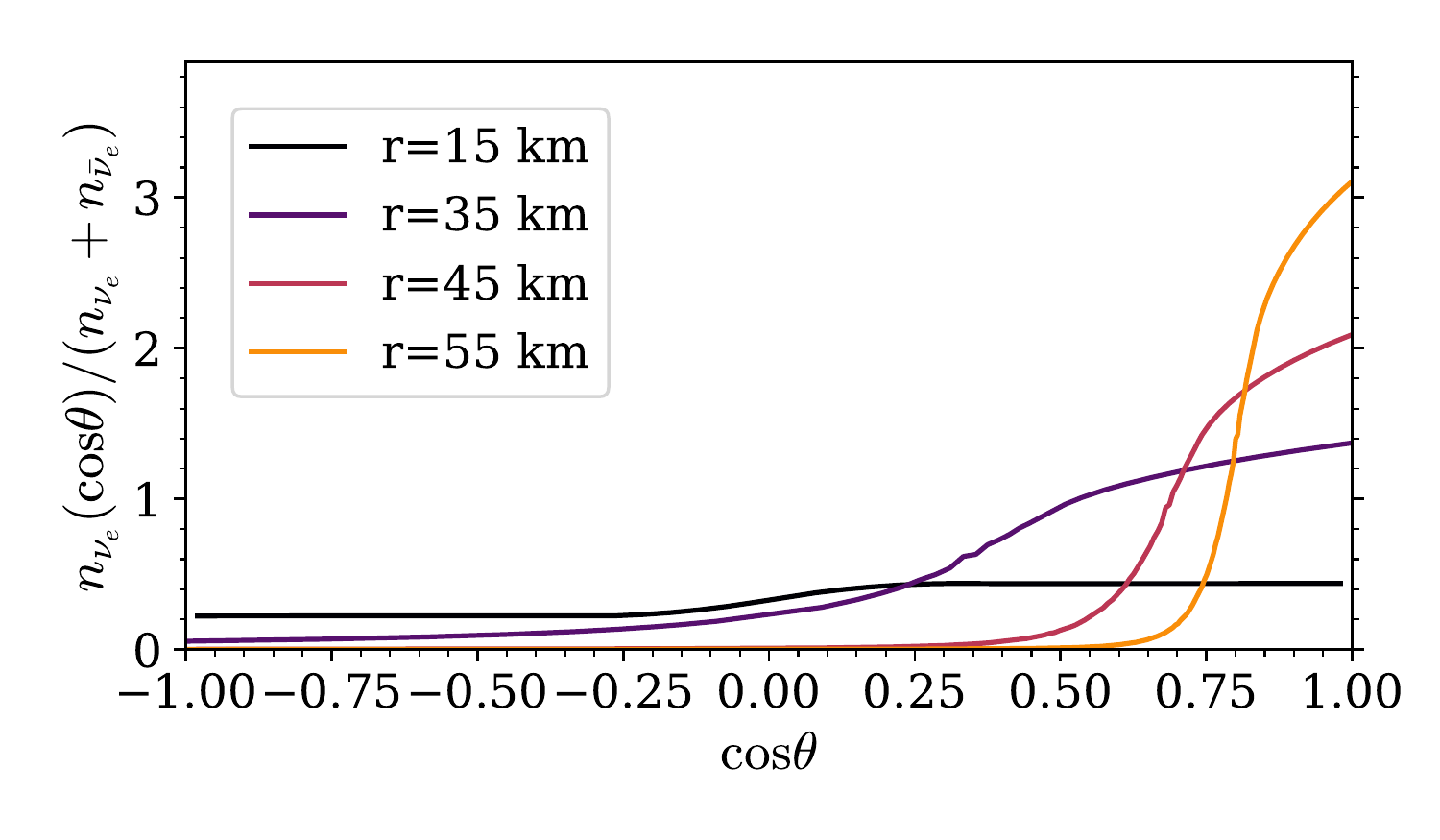}\\
\caption{Illustrative example of the $\nu_e$ angular distribution  as a function of $\cos\theta$  for different radii ($r$) before and after decoupling. The neutrino angular distribution is normalized with respect to the number density of $\nu_e$ after decoupling, at $r=35$~km. The neutrino angular distribution is almost isotropic in $\cos\theta$ in the trapping regime in the SN core; as $r$ increases and neutrinos approach the free-streaming regime, the neutrino angular distribution becomes forward peaked. }
\label{radial_evol}
\end{figure}

Figure~\ref{radial_evol} shows an illustrative example of the resultant $\nu_e$ differential angular distribution,  $n_{\nu_e}(\cos \theta)$,  normalized to the total number density of $\nu_e$ ($n_{\nu_e}$) at $35$~km, i.e, after the neutrino decoupling. The differential angular distributions $n_{\nu_e}(\cos\theta)$, extracted in correspondence of the angular bin centered on $\cos\theta$, has been normalized  to the total number density  $n_{\nu_e} + n_{\bar{\nu}_e}$, defined as,
\begin{eqnarray}
n_{\nu_{e},\bar{\nu}_{e}} = \int_{-1}^{1} n_{\nu_{e},\bar{\nu}_{e}}(\cos\theta) d\cos\theta\ .
\end{eqnarray}

The angular distribution has been plotted at different radii and has been obtained by using the inputs from a 1D hydrodynamical model of a $18.6\,M_\odot$ SN with SFHo nuclear equation of state~\citep{Bollig2016} for $t_{\mathrm{p.b.}} = 0.25$~s, which we will use in Sec.~\ref{sec:snmodel}. One can see that the neutrino angular distribution is almost isotropic in $\cos\theta$ when neutrinos are trapped and becomes forward peaked as neutrinos approach the free streaming regime. In particular,  the $\nu_e$ angular distribution becomes more peaked than the $\bar\nu_e$ as expected by the different interaction rates of electron neutrinos and antineutrinos. We refer the reader to Refs.~\citep{Tamborra:2017ubu,Ott:2008jb,Sarikas:2012vb} for more details on the radial evolution of the flavor-dependent neutrino angular distributions.

\section{Criteria for the appearance of crossings in the electron neutrino lepton number distribution}
\label{sec:crossings}
In this Section, we explore the microphysical conditions leading to the occurrence of ELN crossings. We first focus on a simple toy model in order to pinpoint the main ingredients favoring the development of ELN crossings and present a criterion to predict the  crossing occurrence in the absence of SN hydrodynamical instabilities. The development of ELN crossings in a more realistic SN setup is then investigated by employing inputs from a 1D hydrodynamical simulation of the core collapse. 

\subsection{Toy-model example}
In order to explore what are the conditions leading to the occurrence of ELN crossings, we perform simulations by employing the iterative method described in Sec.~\ref{sec:SNmodel}. For simplicity, we rely on a  simplified setup that does not reproduce the conditions obtained in a realistic SN model; however, we choose to rely on a simplified scheme to explore the main ingredients leading to the development of ELN crossings. We consider a model with  medium temperature  and $\nu_e$ chemical potential not varying with the radius ($T= \mu_{\nu_e} = 10$~MeV).   

In order to characterize the neutrino interaction strength, the total energy-averaged neutrino and antineutrino mean free path is defined as
\begin{equation}
\lambda_{\nu_e, \bar{\nu}_e}^{-1} \simeq \sum_l \left(\frac{1}{\langle\sigma n_{\mathrm{targets}}\rangle}\right)_l^{-1}\ ,
\label{mfptoy}
\end{equation}
where $l$ denotes the various neutrino-matter reaction channels listed in Sec.~\ref{sec:SNmodel}; the neutrino number densities and cross-sections have been modeled as described in Sec.~\ref{sec:SNmodel}, and the Pauli blocking term $F_B$ has been ignored for simplicity. One can easily show that the dominant interaction rates setting the difference between the $\nu_e$ and  $\bar{\nu}_e$ angular distributions, eventually leading to ELN crossings, are the charged current interactions with a subleading contribution coming from the neutral current interactions.  Hence, $\lambda_{\nu_e}/\lambda_{\bar{\nu}_e}$ and the resultant local number density of $\nu_e$ and $\bar\nu_e$ are the most important quantities in setting the relative ratio between the angular distributions of $\nu_e$ and $\bar{\nu}_e$, as also discussed in the Appendix. We arbitrarily assume a constant $\lambda_{\nu_e}/\lambda_{\bar{\nu}_e} \simeq 0.3$ for an electron fraction not varying with the radius ($Y_e = 0.1$).

The baryon density profile is considered to fall exponentially with respect to the radius and distinguish between two cases. A first case involves a shallow baryon density profile (``case A''), 
\begin{eqnarray}
\label{rho1}
\rho_{B, \mathrm{case A}}(r) = 10^{14} \exp\left(0.25(5-r)\right)\ \textrm{gm/cc}\ ,
\end{eqnarray}
and a second case that includes a steeply falling baryon density profile (``case B'')
\begin{eqnarray}
\rho_{B, \mathrm{case B}}(r) = 10^{14}\exp\left(0.5(5-r)\right)\ \textrm{gm/cc}\ .
\end{eqnarray}

\begin{figure}
\includegraphics[width=0.49\textwidth]{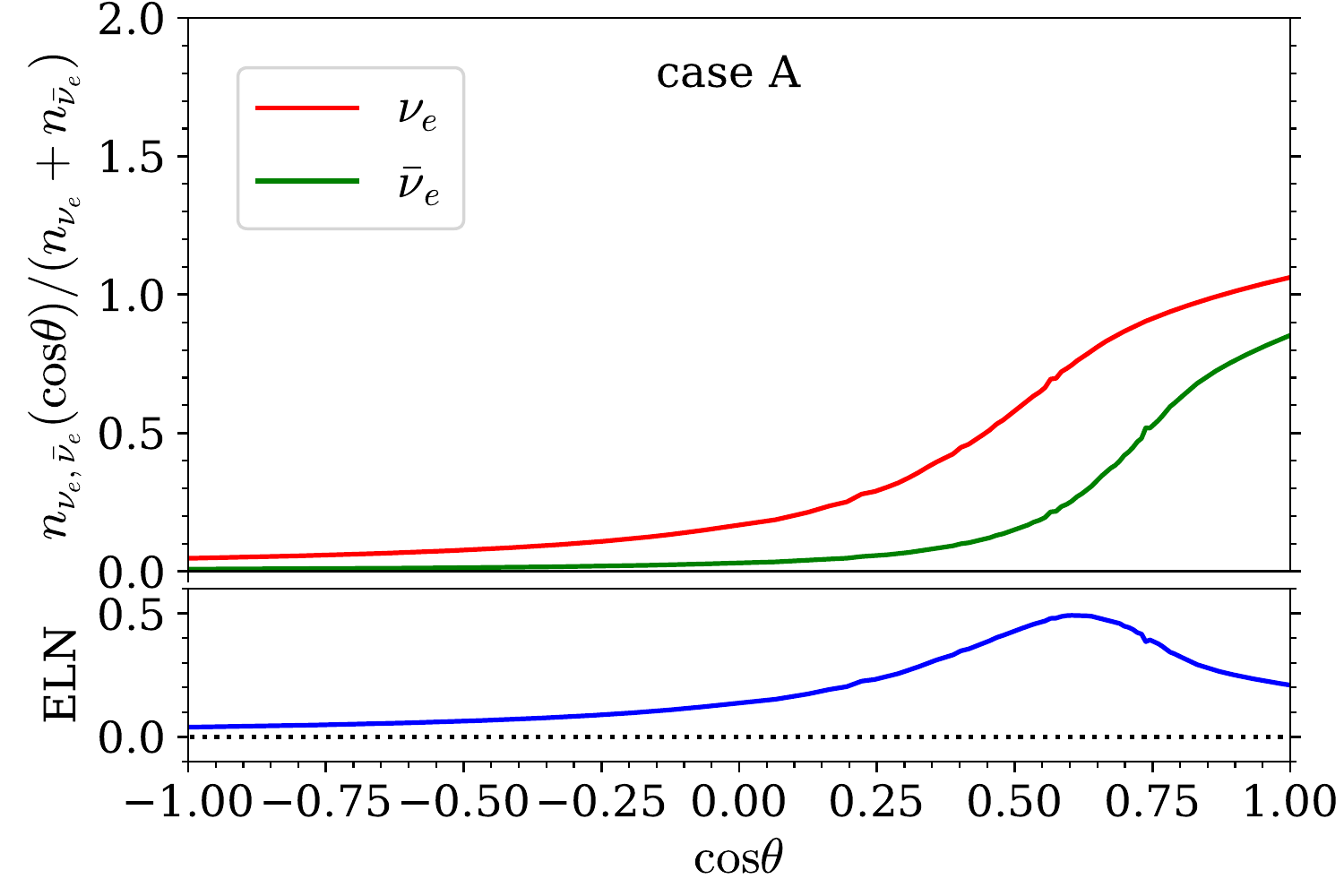}\\
\includegraphics[width=0.49\textwidth]{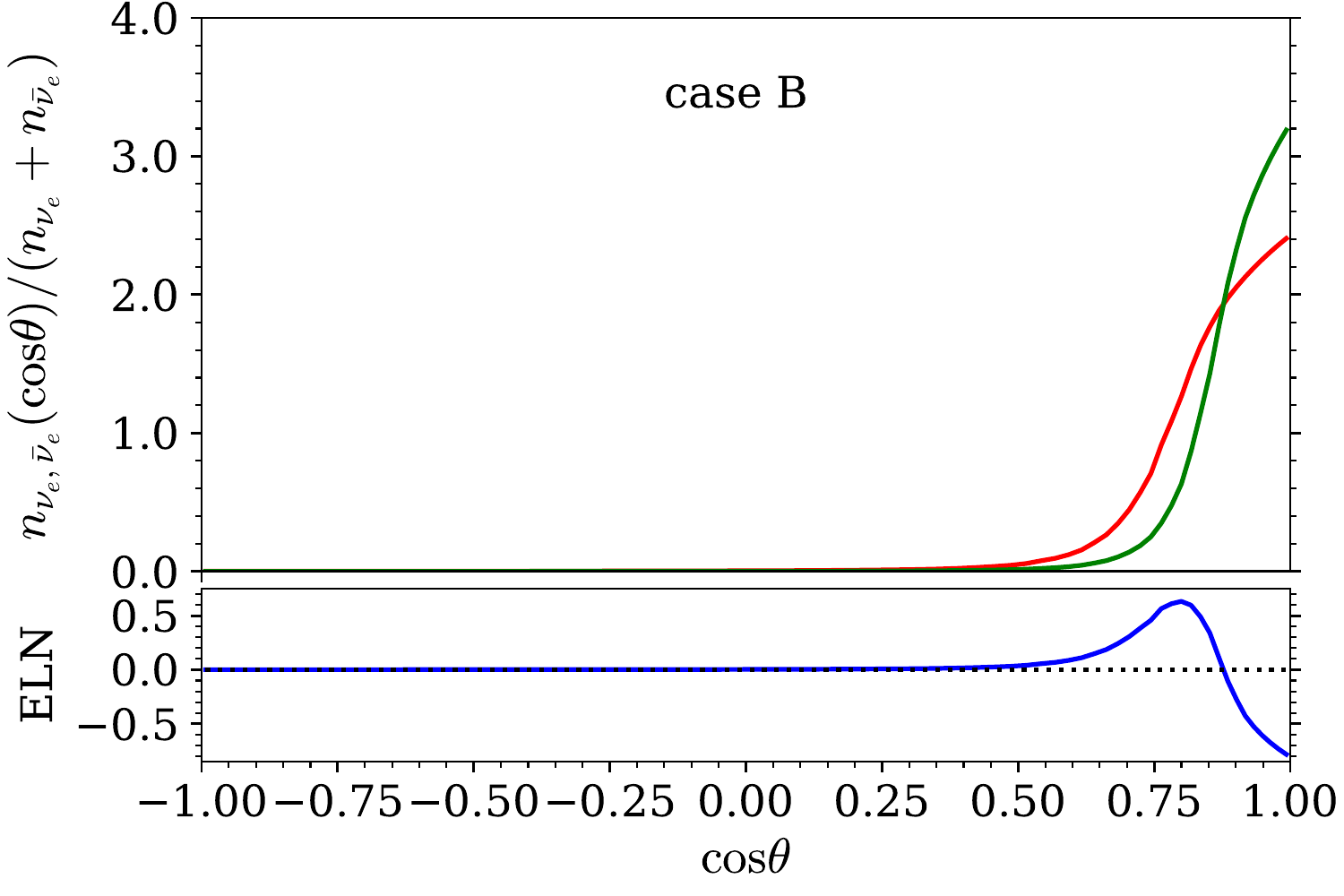}
\caption{{\it Top}: Angular distributions for toy model ``case A'' (see text for details) for $\nu_e$ in red and $\bar{\nu}_e$ in green as a function of $\cos\theta$  normalized  to the total number density of $\nu_e$ and $\bar{\nu}_e$. The angular distributions have been extracted after neutrino decoupling. The shallow baryon density profile ensures that the decoupling radius of $\nu_{e}$ is significantly larger than the one of $\bar{\nu}_{e}$ given the difference in their interaction strength. {\it Bottom}: Angular distributions for toy model ``case B.'' The  rapidly falling baryon density profile implies the close vicinity of the neutrino decoupling radii and the formation of the ELN crossing. The bottom panels for ``case A'' and ``case B'' show the ELN distribution   defined as $[n_{\nu_e}(\cos\theta)-n_{\bar{\nu}_e}(\cos\theta)]/[n_{\nu_e}+n_{\bar{\nu}_e}]$, i.e.~the difference between the red and the green lines). The black dotted line at zero along the $y$-axis is to guide the eye.}
\label{toy1}
\end{figure}
\begin{figure}
\includegraphics[width=0.49\textwidth]{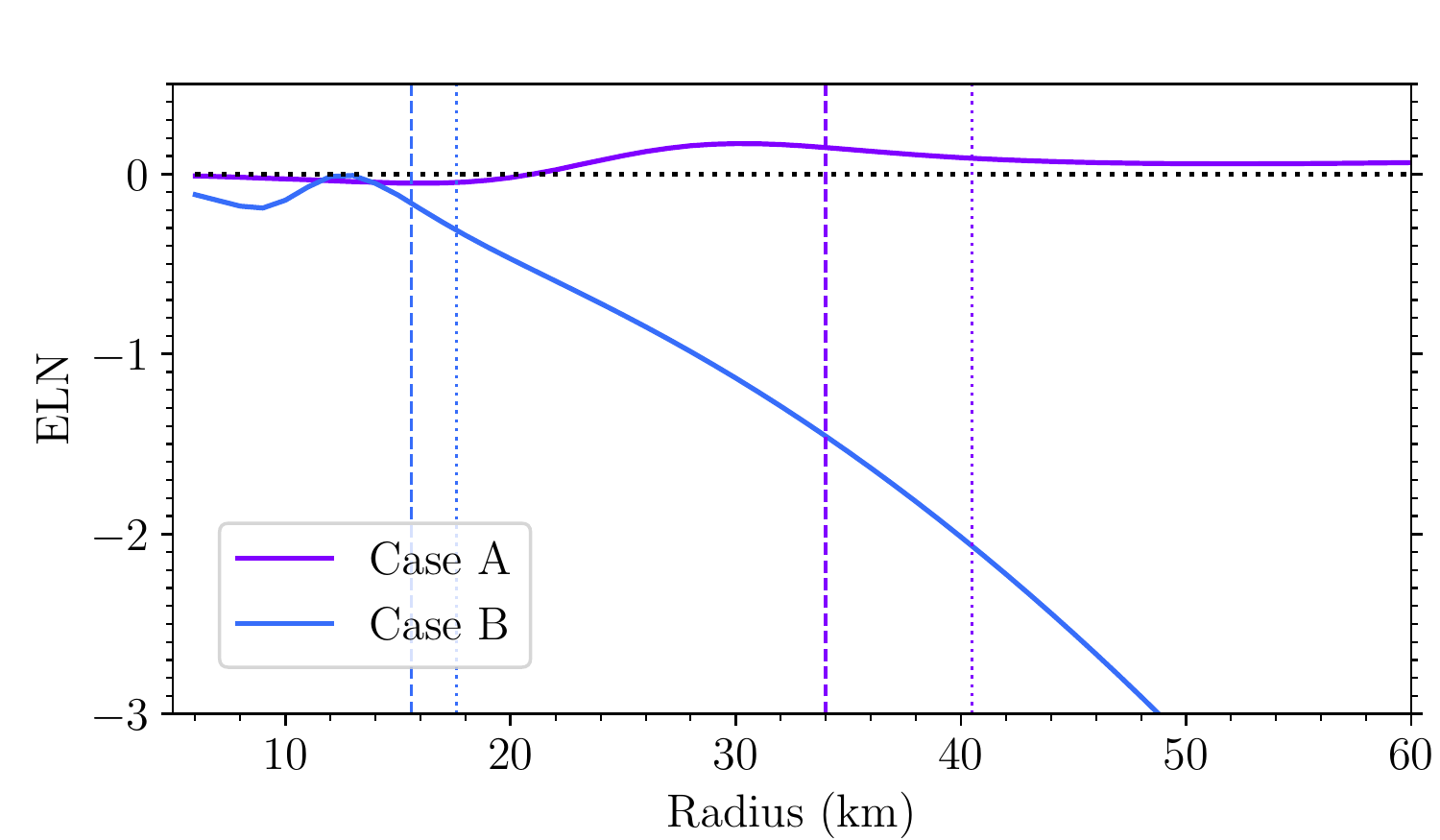}
\includegraphics[width=0.49\textwidth]{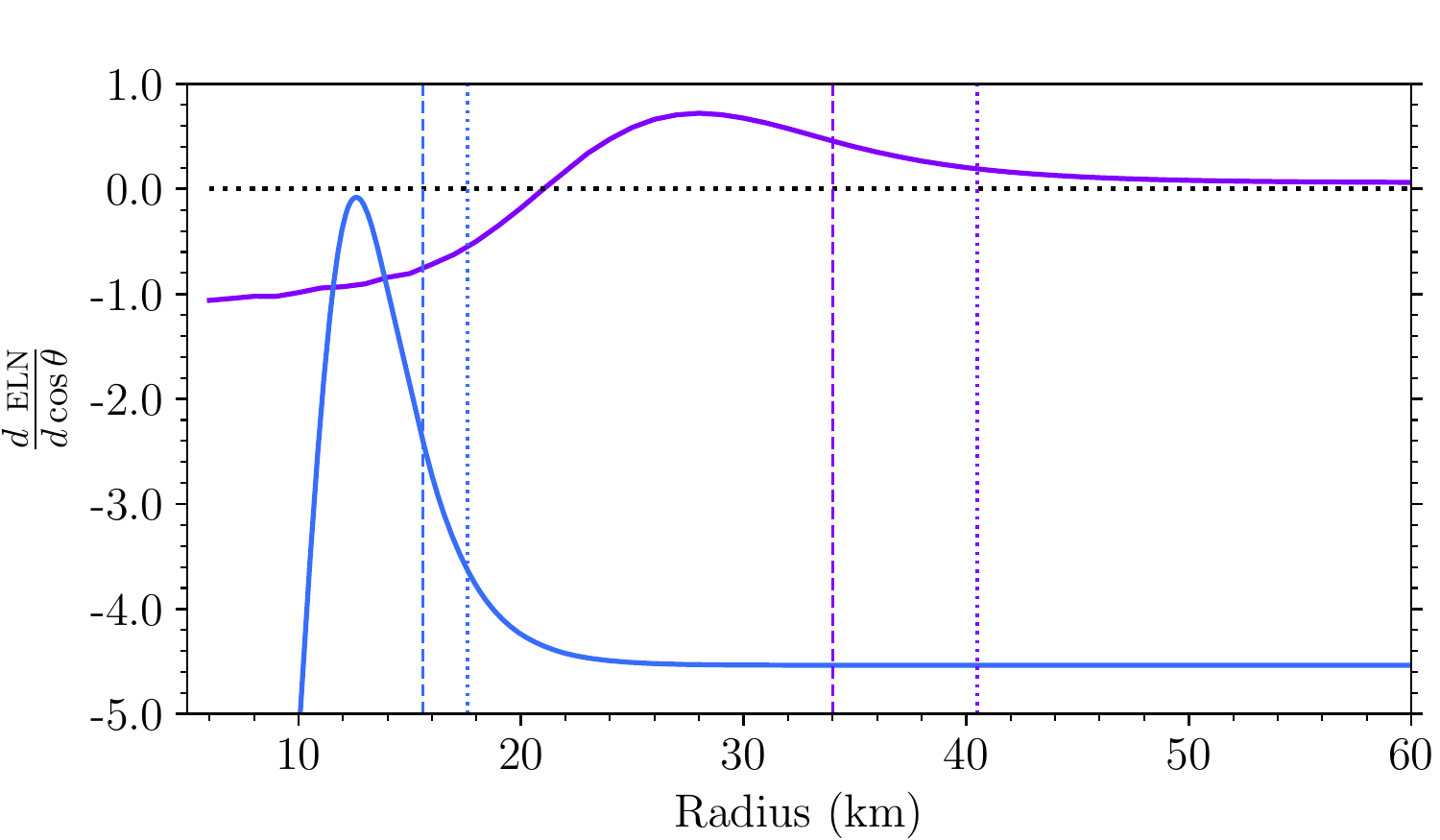}
\caption{{\it Top}: Radial evolution of the ELN,  defined similarly to what done in Fig.~\ref{toy1} but   for the bin containing $\cos\theta=1$ for toy model ``case A''  in violet and ``case B'' in blue.  The radius of onset of free streaming is marked by two vertical lines for each case to guide the eye; the dotted line is for $\nu_e$ and the dashed one for $\bar{\nu}_e$. Neutrinos and antineutrinos decouple at smaller radii for ``case B'' and the onset of free streaming radius of $\nu_e$ is closer to the one of $\bar\nu_e$ for ``case B'' than for ``case A.'' The black dotted line at zero along the $y$-axis is to guide the eye. {\it Bottom}: Radial evolution of  the ELN without integrating over the angular bin. One can see that the the total number of neutrinos is  unchanged at radii larger than the decoupling radius.}
\label{toy1_rad}
\end{figure}

Figure~\ref{toy1} shows the stationary solution obtained for the $\nu_e$ (in red) and $\bar{\nu}_e$ (in green) angular distributions as a function of $\cos\theta$ for ``case A'' on the top and for ``case B'' on the bottom. The differential angular distributions $n(\cos\theta)$ have been normalized  to the total number density  $n_{\nu_e} + n_{\bar{\nu}_e}$. 
Note that the angular distributions of $\nu_x$ and their antineutrinos are also computed, but we do not show them here for simplicity. The plotted angular distributions have been extracted at $r \simeq 40$~km and $r \simeq 30$~km, respectively; i.e., after the neutrino decoupling. One can see that while ``case A'' does not lead to the formation of an ELN crossing, ``case B'' does. For completeness, the bottom panels of ``case A'' and ``case B'' in  Fig.~\ref{toy1} show the resultant ELN distribution ($\nu_e-\bar{\nu}_e$)  and   as a function of $\cos\theta$. 

The top panel of Fig.~\ref{toy1_rad} shows the radial evolution of ELN$(\cos\theta=1)$ for ``case A'' (in violet) and ``case B'' (in blue)\footnote{The ELN has been plotted as indicated in the caption of Fig.~\ref{toy1} but for the angular bin containing $\cos\theta=1$. Note that, since we relied on the local system of coordinates defined by $\theta$, the width of the bin that includes $\cos\theta=1$ depends on the radius. This is not true in the global system of coordinates defined by $\theta_0$ and the number of particles remains constant after decoupling.}. For each case, the onset of the free-streaming regime of $\nu_e$ and $\bar\nu_e$  is marked by two vertical lines to guide the eye;  since we only need an approximate estimation of the region where the neutrinos start to stream freely and are fully decoupled from matter, we define the radius of onset for the free-streaming regime as the radius  where the forward flux ($\cos\theta>0$) is 10 times larger than the backward flux ($\cos\theta<0$). One can see that the free-streaming regime is reached at larger $r$ for ``case A,'' as expected. Moreover, the onset radius of free-steaming for $\nu_e$ is always  larger than the one of $\bar\nu_e$ given the difference in the interaction strengths. The two radii are separated by a larger distance in ``case A.'' The bottom panel of Fig.~\ref{toy1_rad} shows the radial evolution of the ELN without integrating over the $\cos\theta$ bin. As expected, the total number of $\nu_e$ and $\bar{\nu}_e$ is constant at radii larger than the decoupling one.

By looking at Figs.~\ref{toy1}  and \ref{toy1_rad}, one can see that ELN crossings originate in the proximity of the neutrino free-streaming region in ``case B.'' Moreover, because of the different baryon profiles employed  in ``case A'' and ``case B,'' the decoupling regions of $\nu_e$ and $\bar{\nu}_e$ occur in very different spatial regions for ``case A'' while they are close to each other for ``case B.'' This implies that  the  $\nu_e$ and $\bar{\nu}_e$ angular distributions are more similar to each other in the proximity of the decoupling region in ``case B'' than in ``case A.'' This is also proved by the fact that the total local number densities of $\nu_e$ and $\bar{\nu}_e$ are more similar to each other in the proximity of the decoupling region in ``case B,'' as shown in Table~\ref{table1}. 
{\def\arraystretch{1.35}
\begin{table}
\centering
\caption{Ratio of the electron neutrino and antineutrino number densities ($n_{\nu_e}/n_{\bar{\nu}_e}$) at the radius of the onset of free streaming of $\nu_e$ for ``case A'' and ``case B,'' see text for details.}
\begin{tabular}{c|c}
		&$n_{\nu_e}/n_{\bar{\nu}_e}$							\\\hline\hline
case A	&$2.02$			\\
case B	&$1.29$	
\end{tabular}
\label{table1}
\vspace*{0.06in}
\end{table}}

This toy-model example shows that ELN crossings can develop only in the proximity of the neutrino decoupling region. Moreover,  the decoupling of $\nu_{e}$ should occur in a radial region close to where the one  of  $\bar{\nu}_{e}$ happens and  their local number densities should be comparable. If those conditions are fulfilled, ELN crossings are likely to develop.

\subsection{Supernova model example}
\label{sec:snmodel}
We now extend our findings to a more complex case involving the radial  dependence of the main SN quantities. We base our estimations on the inputs from a  1D hydrodynamical model of a $18.6\,M_\odot$ SN with SFHo nuclear equation of state and gravitational mass of $1.4\,M_\odot$~\citep{Bollig2016} that we adopt as benchmark case, and select post-bounce time snapshots representative of the different SN phases. Note that the hydrodynamical simulation does not provide the neutrino angular distributions that are instead estimated iteratively through our stationary and spherically symmetric model.

\begin{figure}
\includegraphics[width=0.49\textwidth]{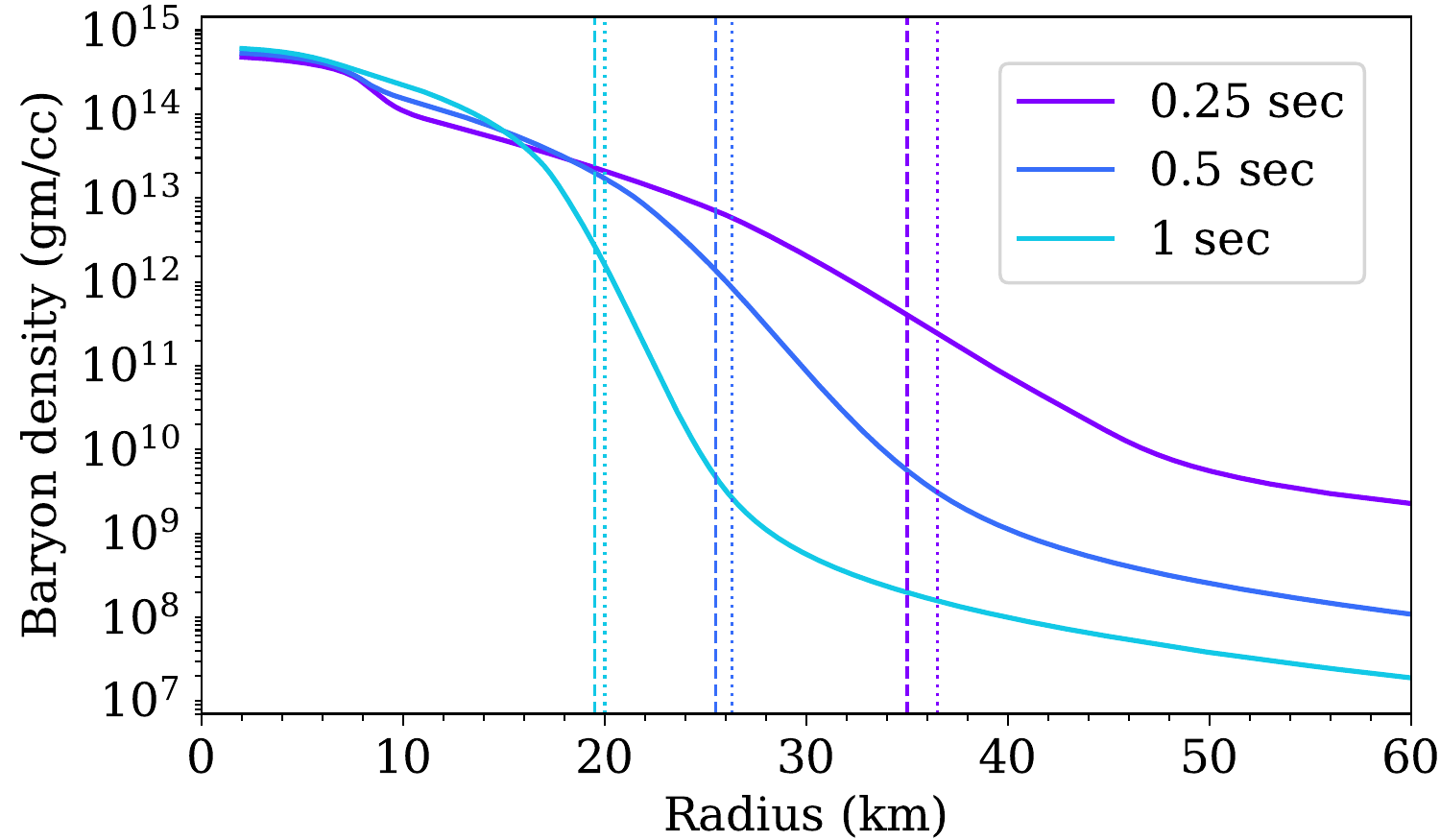}\\
\hspace{100in}\includegraphics[width=0.495\textwidth]{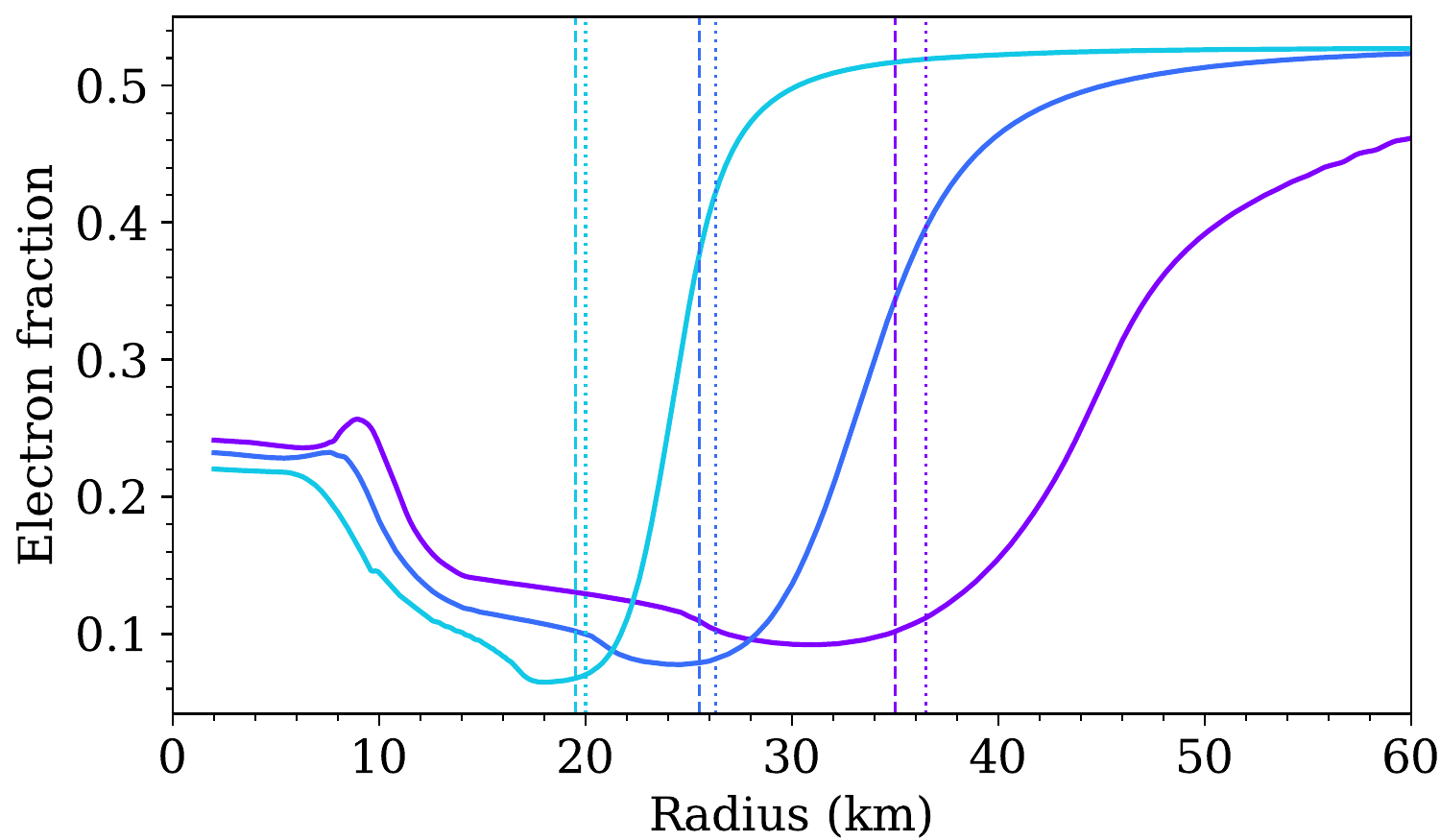}\\
\hspace{15in}\includegraphics[width=0.495\textwidth]{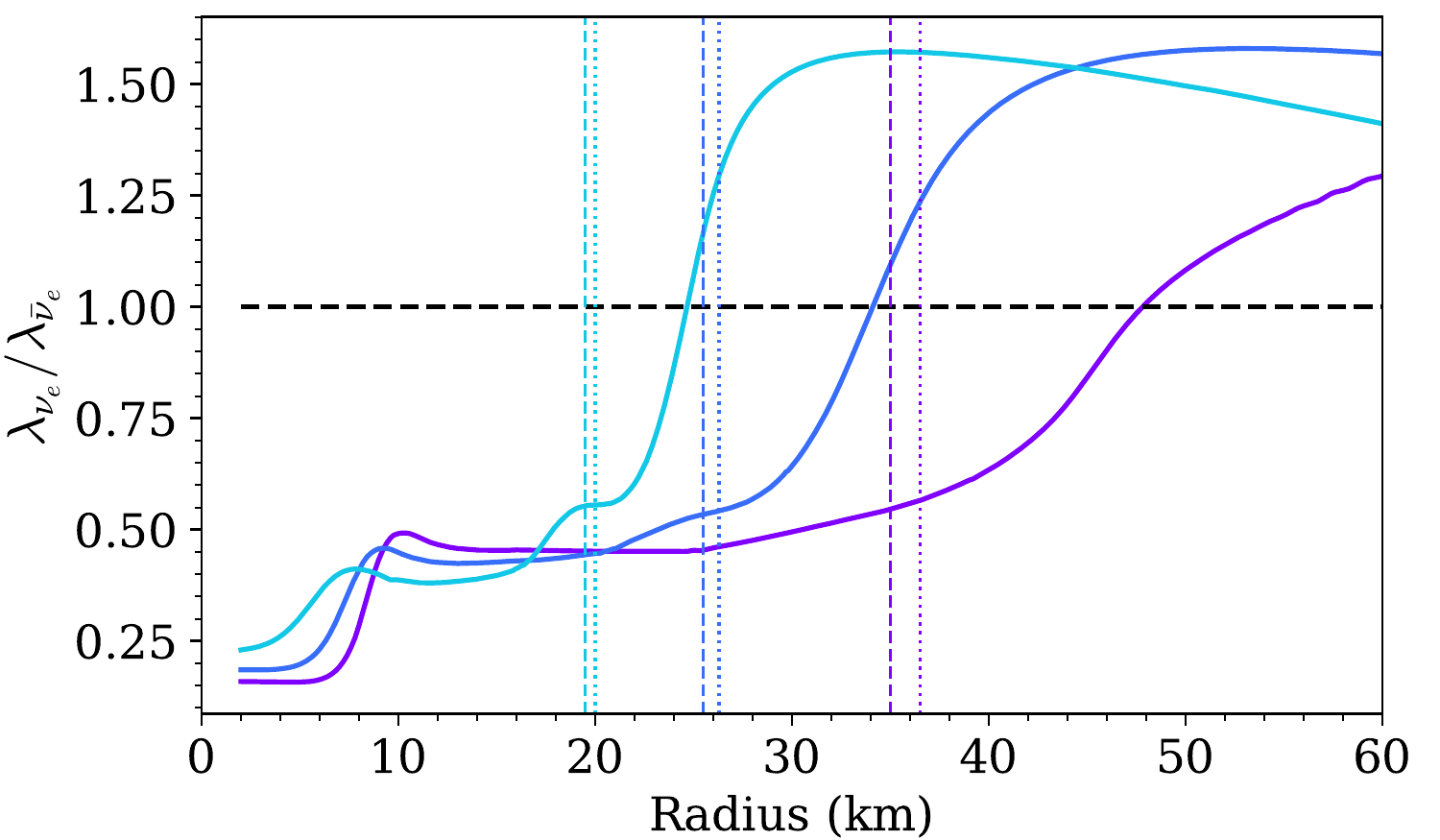}
\caption{{\it Top}: Baryon density as a function of radius for $t_{\mathrm{p.b.}} = 0.25, 0.5$ and $1$~s, in violet, blue and cyan respectively, for the $18.6\,M_\odot$ SN model adopted as benchmark case in this work.  The radii of onset of free streaming are marked by two vertical lines, the  dotted one is for $\nu_e$ and the  dashed one for $\bar{\nu}_e$, to guide the eye.  In the region of decoupling the baryon density falls off much more rapidly as $t_{\mathrm{p.b.}}$ increases. {\it Middle}: Electron abundance  as a function of radius for $t_{\mathrm{p.b.}} = 0.25, 0.5$ and $1$~s. $Y_e$ is smaller in the proximity of the decoupling radius for larger $t_{\mathrm{p.b.}}$. {\it Bottom}: Ratio of mean free paths of $\nu_{e}$ and $\bar{\nu}_{e}$. It should be noted that the mean free path of $\bar{\nu}_{e}$ is larger than that of ${\nu}_{e}$ before decoupling.}
\label{density}
\end{figure}

The top panel of Fig.~\ref{density} shows  the baryon density as a function of the radius for three different post-bounce times $t_{\mathrm{p.b.}} = 0.25, 0.5$ and $1$~s, in violet, blue and cyan respectively. The radii of the onset of free-streaming of $\nu_e$ and $\bar\nu_e$ are marked through the vertical lines to guide the eye. One can see that, as $t_{\mathrm{p.b.}}$ increases, neutrinos start to free-stream at smaller radii, closer to the SN core. Moreover, as time progresses, the baryon density profile becomes steeper in the proximity of the region of the onset of free streaming. Another interesting aspect is that, for fixed $t_{\mathrm{p.b.}}$, the free-streaming radii of $\nu_e$ and $\bar{\nu}_e$ differ from each other at earlier post-bounce times and become similar to each other at later times. 

For earlier times, the electron fraction $Y_e$ is larger at radii smaller  than the free-streaming one  as shown in the middle panel of Fig.~\ref{density}. 
The bottom panel of Fig.~\ref{density}  shows  the ratio of the mean free paths $\lambda_{\nu_e}/\lambda_{\bar\nu_e}$ as a function of the radius; in the proximity of the free-streaming radius, $\lambda_{\nu_e}/\lambda_{\bar\nu_e} \le 1$. The ratio between the $\nu_e$ and $\bar\nu_e$ number densities at the radius of onset of free streaming is reported in Table~\ref{table2} for the three studied $t_{\mathrm{p.b.}}$. One can see that $n_{\nu_e}/n_{\bar{\nu}_e} \rightarrow 1$ as $t_{\mathrm{p.b.}}$ increases.
{\def\arraystretch{1.35}
\begin{table}
\centering
\caption{Ratio of the electron neutrino and antineutrino number densities ($n_{\nu_e}/n_{\bar{\nu}_e}$) at the radius of the onset of free streaming of $\nu_e$  for three different post-bounce times $t_{\mathrm{p.b.}} = 0.25, 0.5$ and $1$~s  of our benchmark SN model.}
\begin{tabular}{c|c}
		&$n_{\nu_e}/n_{\bar{\nu}_e}$							\\\hline\hline
$t_{\mathrm{p.b.}} = 0.25$~s	&$1.30$			\\
$t_{\mathrm{p.b.}} = 0.5$~s	&$1.14$	 \\
$t_{\mathrm{p.b.}} = 1$~s	&$1.07$	 
\end{tabular}
\label{table2}
\vspace*{0.06in}
\end{table}}

\begin{figure}
\includegraphics[width=0.49\textwidth]{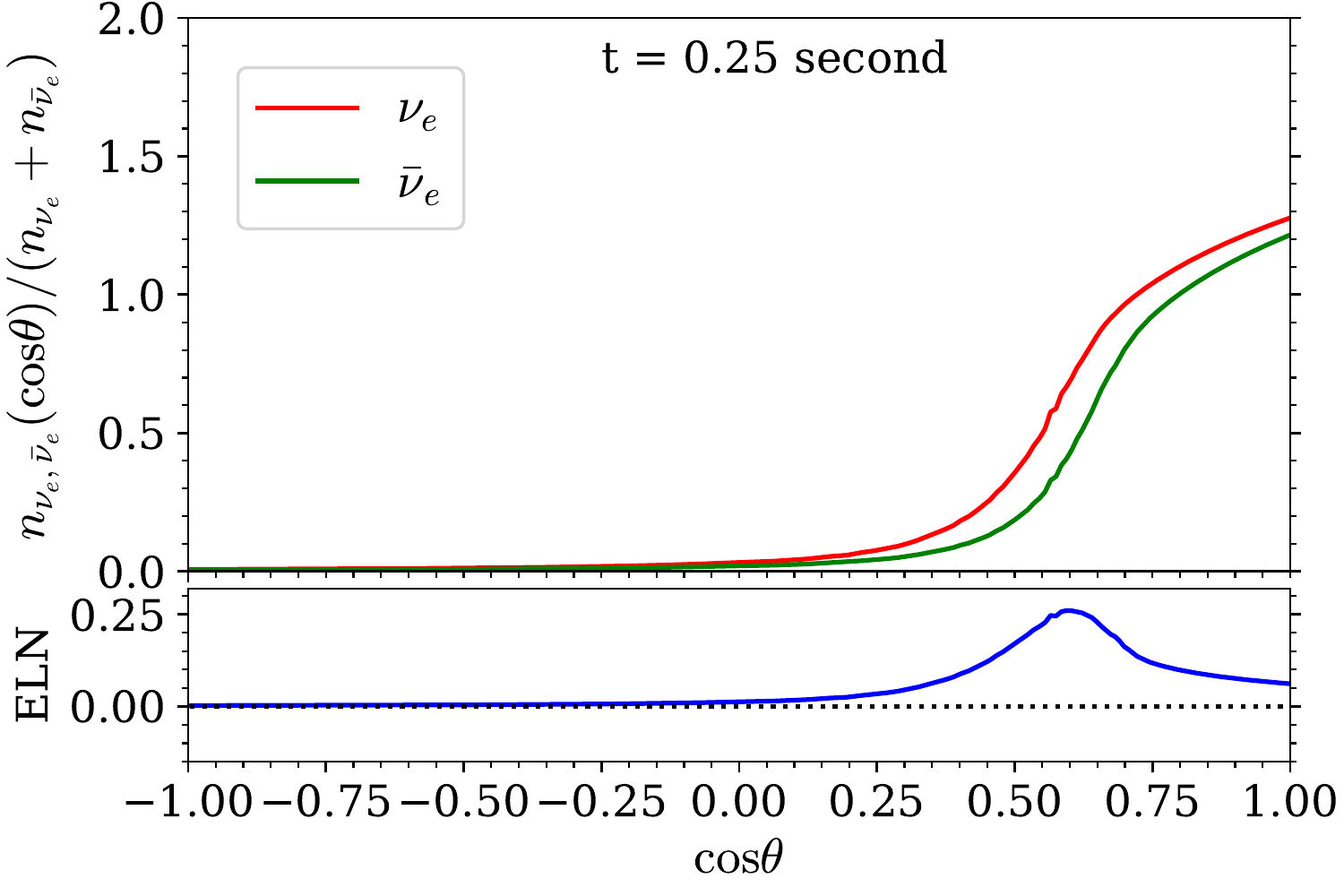}\\
\includegraphics[width=0.49\textwidth]{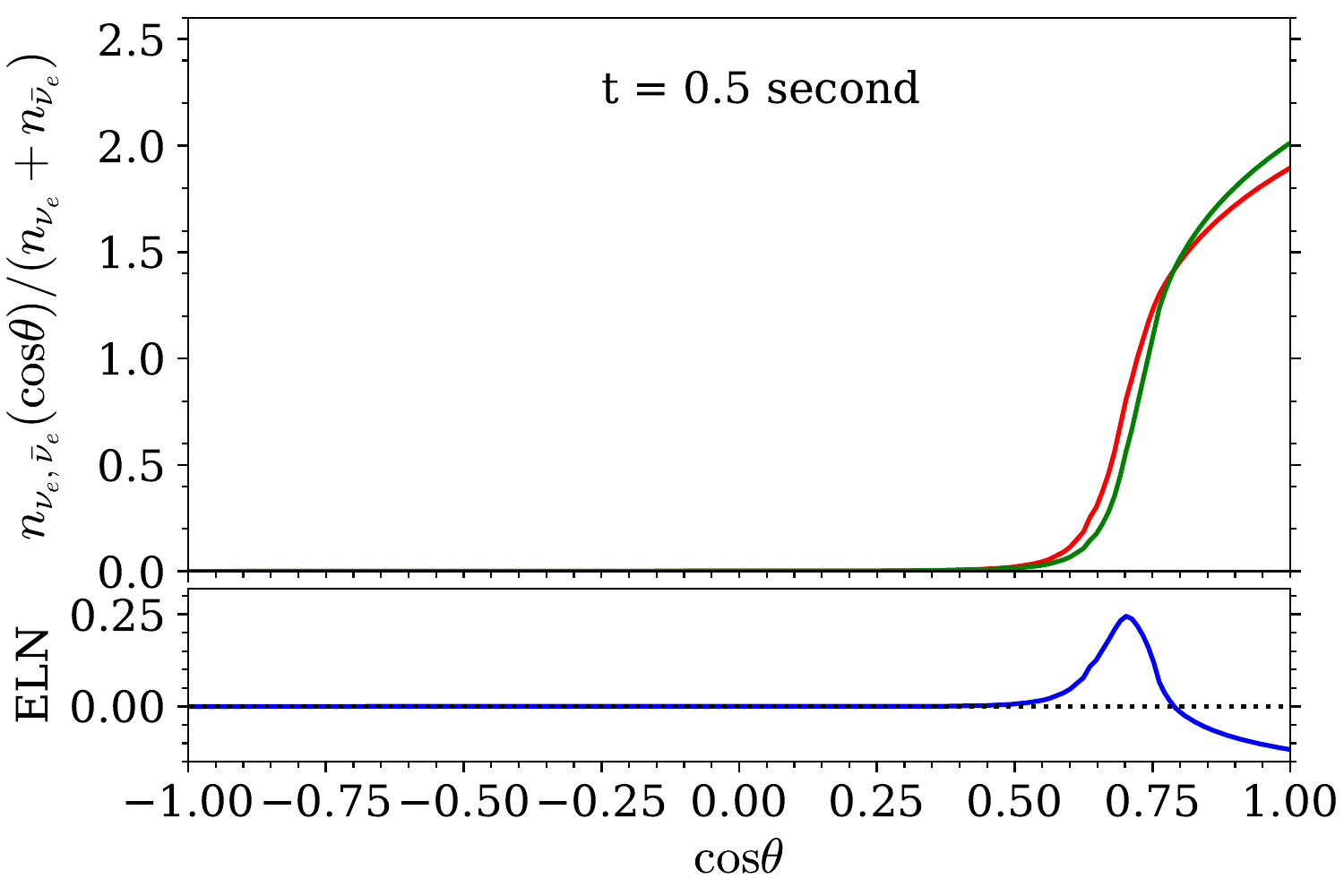}\\
\includegraphics[width=0.49\textwidth]{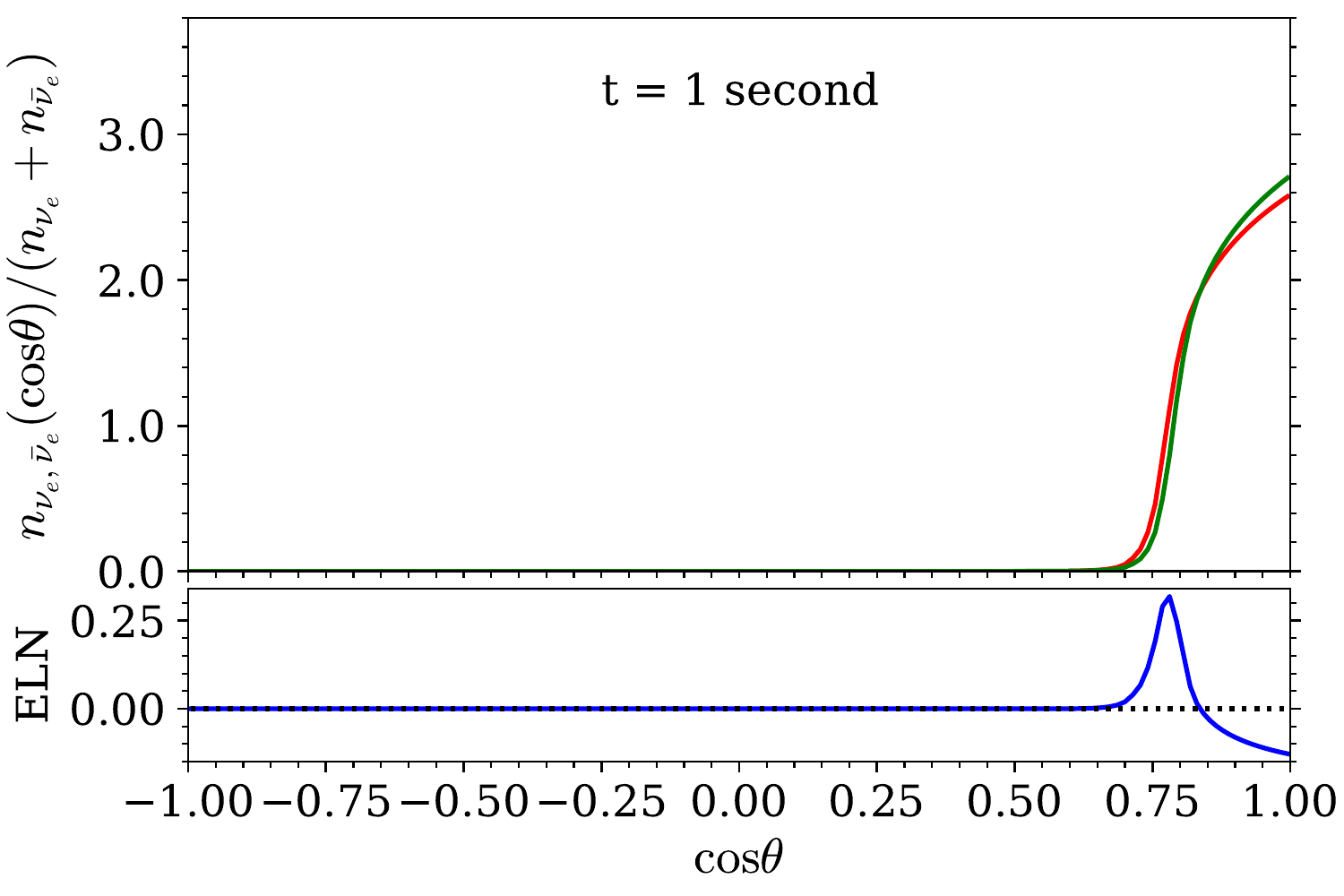}
\caption{{\it Top}: Angular distributions of $\nu_{e}$ in red and $\bar{\nu}_{e}$ in green,  each normalized to the total number density of $\nu_e$ and $\bar{\nu}_e$,  for $t = 0.25$~s as a function of $\cos\theta$ for our benchmark SN model. The angular distributions have been extracted at 40~km, i.e.~after decoupling. The lower panel shows the angular distribution of the ELN. The excess in the total number of $\nu_{e}$ prevents an ELN crossing. {\it Middle}:  Angular distributions of $\nu_{e}$ and $\bar{\nu}_{e}$ for $t = 0.5$~s at 35~km. An ELN crossing occurs  as the decoupling regions of $\nu_e$ and $\bar{\nu}_e$ become closer to each other and their number densities are comparable.  {\it Bottom}: Angular distributions of $\nu_{e}$ and $\bar{\nu}_{e}$ for $t = 1$~s at 30~km. The ELN becomes negative 
 in the forward direction implying that a crossing occurs.  The black dotted line at zero along the $y$-axis is to guide the eye.
}
\label{real1}
\end{figure}

Figure~\ref{real1}  shows the resultant angular distributions for $\nu_e$ and $\bar{\nu}_e$,  each normalized to the total number density of $\nu_e$ and $\bar{\nu}_e$, as a function of $\cos\theta$ for $t_{\mathrm{p.b.}} = 0.25, 0.5$ and $1$~s from top to bottom, respectively. Those angular distributions have been obtained  with our iterative method by employing the SN inputs shown in Fig.~\ref{density}, together with the radial profiles of the chemical potentials for neutrinos and nucleons and the medium temperature extracted from the hydrodynamical simulation. We plot the angular distributions  at an arbitrary radius  of $40$~km, i.e.~after the neutrino decoupling occurred. 
 
 For $t_{\mathrm{p.b.}} = 0.25$~s, no ELN crossing is found as the number density of $\nu_{e}$ remains larger than the one of $\bar{\nu}_{e}$ due to the shallow baryon density profile. At late times ($t_{\mathrm{p.b.}} \gtrsim 0.5$~s), the baryon density profile becomes steeper; as a result the free-streaming radii of $\nu_{e}$ and $\bar{\nu}_{e}$ become closer and the number densities of $\nu_e$ and $\bar{\nu}_e$ become comparable. Hence, ELN crossings are favored. Note that we investigated the appearance of crossings by using inputs from 5 time snapshots between $0.125$~s and $1$~s in our SN model [$t_{\mathrm{p.b.}} = (0.125, 0.25, 0.5, 0.75, 1)$~s], but we only show results for three of them for simplicity. The first post-bounce time for which we find ELN crossings among the analyzed ones is $0.5$~s. ELN crossings appear for any $t_{\mathrm{p.b.}} \gtrsim 0.5$~s of the ones studied.

\begin{figure}
\includegraphics[width=0.49\textwidth]{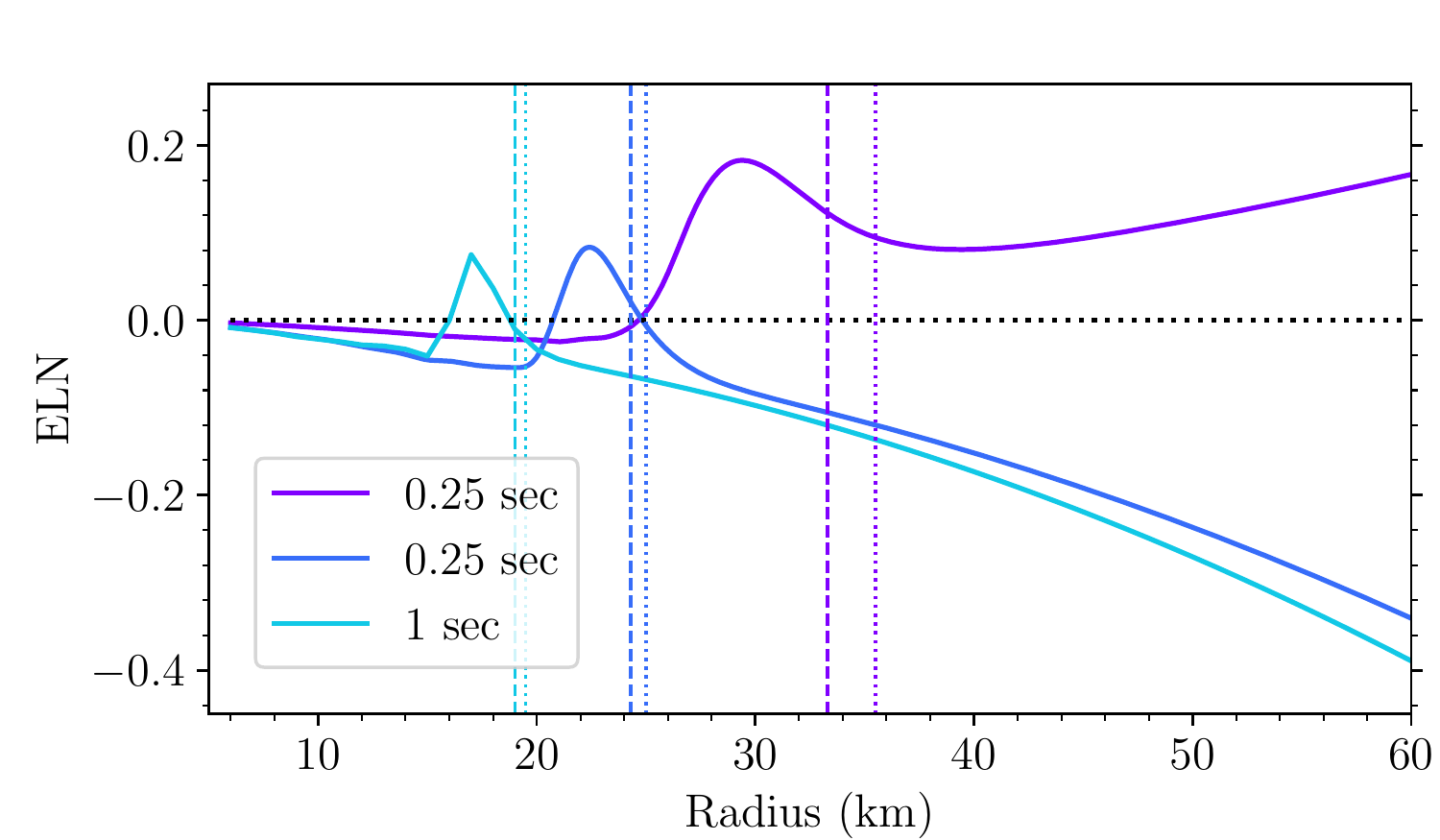}
\includegraphics[width=0.49\textwidth]{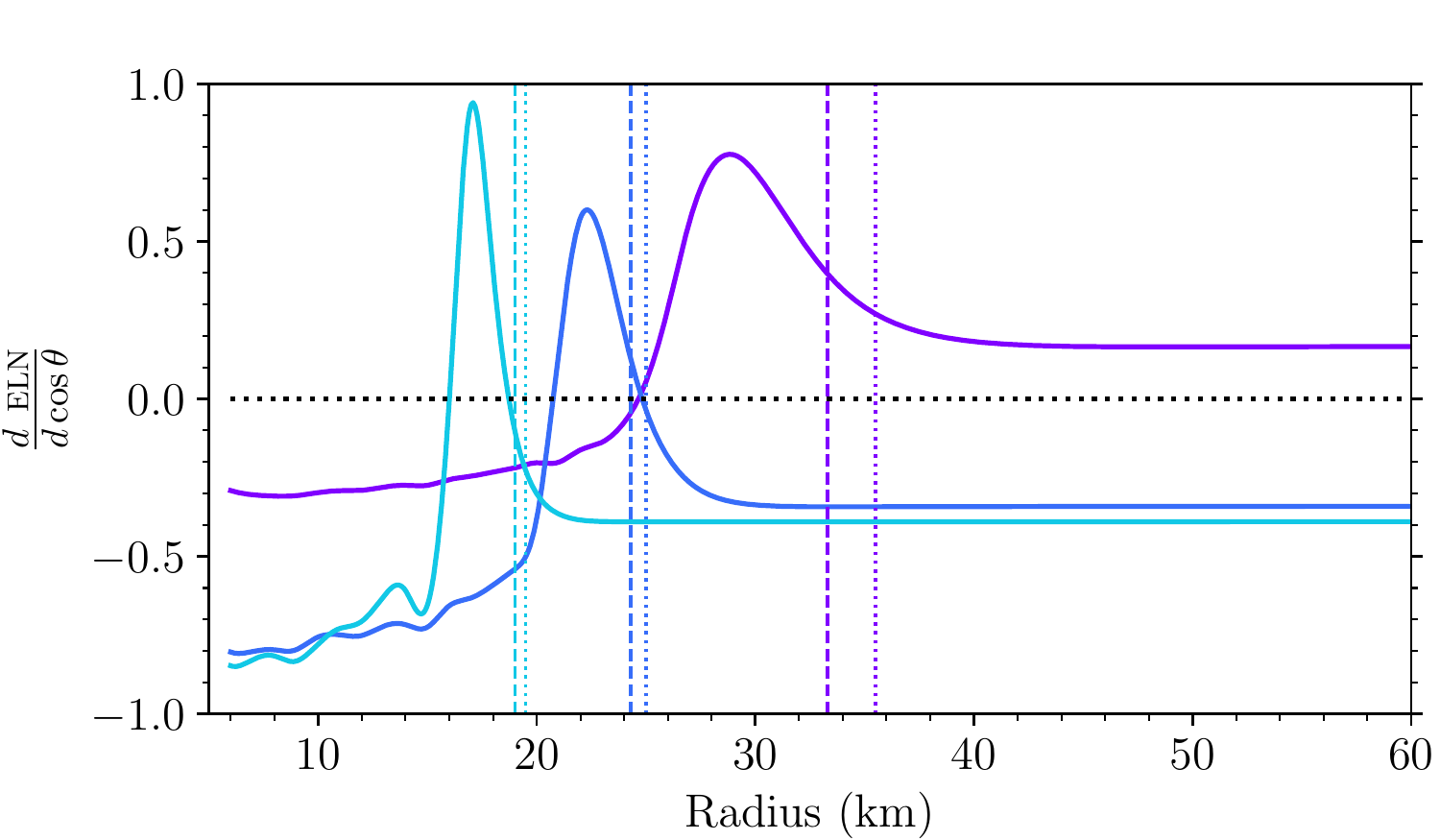}
\caption{{\it Top}: Radial evolution of ELN  for angular bin containing $\cos\theta = 1$, as a function of the radius  normalized to the total number density of $\nu_e$ and $\bar{\nu}_e$, for $t_{\mathrm{p.b.}} = 0.25, 0.5$ and $1$~s, in violet, blue, and cyan respectively. For $t_{\mathrm{p.b.}} = 0.5$ and $1$~s,  ELN crossings appear in the proximity of the neutrino free-streaming region, i.e., ELN$(\cos\theta = 1)$ changes sign.  The black dotted line at zero along the $y$-axis is to guide the eye.
{\it Bottom:} Radial evolution of ELN without integrating over the angular bin; the total number of neutrinos is  unchanged at radii larger than the decoupling radius.}

\label{real1_rad}
\end{figure}

The top panel of Fig.~\ref{real1_rad} shows the radial evolution of ELN$(\cos\theta = 1)$ for $t_{\mathrm{p.b.}} = 0.25, 0.5$ and $1$~s. As discussed for the toy-model, ELN crossings only occur  in the proximity of the neutrino free-streaming region. In  correspondence of the appearance of crossings, we expect that ELN$(\cos\theta = 1)$ changes sign; this is what we find for $t_{\mathrm{p.b.}} = 0.5$ and $1$~s.
The total number of neutrinos and antineutrinos remains unchanged after decoupling as shown in the bottom panel of Fig.~\ref{real1_rad}.

In conclusion, our stationary and spherically symmetric SN model strongly suggests that  ELN crossings can only occur within the spatial regions where neutrinos and antineutrinos decouple and start to free stream. Moreover, a steep drop of the baryon density profile, as typical of the late SN stages, together with comparable number densities of $\nu_e$ and $\bar{\nu}_e$ and  $\lambda_{\nu_{e}}/\lambda_{\bar{\nu}_{e}} \le 1$ in the decoupling region, would favor the development of angular distributions that are similar to each other; as a consequence, ELN crossings develop given the different interaction rates of  $\nu_e$ and $\bar{\nu}_e$.  Our results confirm the findings of Refs.~\citep{Tamborra:2017ubu,Azari:2019jvr} where only the early SN stages were analyzed and no crossing was found. 
Notably,  hydrodynamical simulations investigating the behavior of the neutrino angular distributions presented in the literature do not exhibit successful explosions or have been artificially exploded; as a consequence, none of them investigates the behavior of the neutrino angular distributions in the late accretion and cooling phases. Our model is the first to predict the occurrence of crossings at later post-bounce times, when the baryon density profile becomes steeper.

Our stationary and spherically symmetric SN model does not include any macroscopic asymmetry eventually induced by  hydrodynamical instabilities. 
The occurrence of the LESA instability has been pinpointed as a possible favorable condition leading to ELN crossings~\citep{Izaguirre:2016gsx,Dasgupta:2016dbv}. Our stationary SN model hints that,  in the SN angular regions of net $\nu_e$ (or $\bar{\nu}_e$) excess,  our criteria should still hold.  Hence, we expect that ELN crossings should occur only across  the SN angular regions where the ELN changes sign because of LESA (see, e.g., Fig.~1 of \cite{Tamborra:2014aua}). On the other hand, if the LESA instability is sustained until the late SN phases, then ELN crossings may develop even in regions of net $\nu_e$ (or $\bar{\nu}_e$) excess, as predicted for the cases with $t_{\mathrm{p.b.}} \ge 0.5$~s in our SN model. 
We stress, however, that our model cannot fully test the conditions leading to crossings in the presence of the LESA instability self-consistently, since this would require a break of the spherical symmetry. Therefore, our conjectures remain to be tested thorough self-consistent 3D hydrodynamical simulations and will be subject of further work.

\section{Dynamical generation of crossings in the electron neutrino lepton number distribution}
\label{sec:oscillations}
We have seen in the previous Sections that the presence of ELN crossings, or lack thereof, is determined by the radial profile of $\lambda_{\nu_{e}}/\lambda_{\bar{\nu}_{e}}$, the baryon density profile, as well as  $n_{\nu_e}/n_{\bar{\nu}_e}$, which is related to the former two, in the trapping region.  However, as discussed in Sec.~\ref{sec:crossings}, the presence of ELN crossings is not only determined by the local conditions, but it indirectly feels the effect of distant regions in the SN through collisions.  In this Section, we intend to explore whether ELN crossings can be generated dynamically from flavor instabilities  occurring in the SN  because of local fluctuations~\citep{Abbar:2015fwa,Dasgupta:2015iia,Capozzi:2016oyk}. According to this scenario, the neutrino angular distributions may be modified dynamically; as a consequence, ELN crossings may form and, in turn, they  could foster the growth of fast flavor instabilities.

It should be noted that the neutrino interaction rate in matter is dominated by the neutrino-nucleon cross-section, and the ratio $\lambda_{\nu_{e}}/\lambda_{\bar{\nu}_{e}}$  cannot be changed significantly by the neutrino conversions on a global scale. In the deep interior of a SN, the effective average  energy  of $\nu_{e}$ is larger than that of $\bar{\nu}_{e}$ due to the non-null chemical potential. This contributes to a larger $\nu_{e}$ interaction rate. Neutrino conversions could in principle reduce the average energy of $\nu_{e}$  and bring it closer the one of $\bar{\nu}_{e}$.

We take the inputs of the SN snapshot at $t_{\mathrm{p.b.}} = 0.25$~s  used in Sec.~\ref{sec:snmodel} as benchmark case. This case did not exhibit ELN crossings in the absence of pre-existing flavor conversions. We then  impose that  at a certain radius $r_\star$, with $r_\star \in [r_{\mathrm{min}} = 5~\mathrm{km}, r_{\mathrm{max}} = 60~\mathrm{km}]$, flavor conversions are triggered possibly leading to flavor decoherence;  the latter is one of the most extreme scenarios that one could expect. This effect can be mimicked by assuming that  $\lambda_{\nu_{e}}/\lambda_{\bar{\nu}_{e}} \rightarrow 1$ for $r \ge r_\star$.  We then change $r_\star \in [r_{\mathrm{min}}, r_{\mathrm{max}}]$  in order to test whether the radius of the onset of flavor conversions would affect the dynamical development of ELN crossings. 

In none of the studied cases, we find a significant modification of the ELN evolution and ELN crossings  do not develop  within our simplified SN setup. It is worth noticing that the specific choice of the numerical values adopted for $r_{\mathrm{min}}$ and $r_{\mathrm{max}}$ does not affect our conclusions as long as both radii are chosen  to be far away from the decoupling region.

We then conclude that, in our model, the neutrino angular distributions cannot be dynamically modified by flavor instabilities due to local fluctuation and induce ELN crossings dynamically.
Although a more thorough analysis of the problem is required, we find that the presence of fast oscillations at radii much larger than the decoupling radius, cannot significantly change conditions in the region of decoupling to facilitate the dynamical generation of ELN crossings in a geometry that is spherically symmetric. 

\section{Discussion and Conclusions}
\label{sec:conclusions}
The development of crossings between the angular distributions of $\nu_e$ and $\bar{\nu}_e$ (ELN crossings) is of relevance because it can  possibly lead to  fast pairwise conversions of neutrinos deep in the SN core with major consequences on the SN physics. It is vital to gain a qualitative understanding of this phenomenon. 
In this paper, using a simple yet insightful technique, we have  qualitatively addressed this question focusing on the microphysics of neutrino-matter collisions in the SN core. 

The highly non-linear nature of the neutrino flavor evolution along with the feedback on the flavor dynamics coming from  neutrino-matter collisions makes a general self-consistent analysis impossible within current means. However, we have aimed to provide a rule of thumb for the occurrence of ELN crossings. 
To that purpose, we have constructed a simplified stationary and spherically symmetric SN model that takes into account the physics of collisions through an iterative approach, but neglects any asymmetry and further complications coming from the SN hydrodynamical instabilities. 
It is should be noted that  our assumption of spherical symmetry may be substantially broken  either due to SN hydrodynamics~\citep{Tamborra:2014hga,Janka:2016fox}, or because the nature of the neutrino flavor evolution~\citep{Duan:2014gfa,Mirizzi:2015fva,Abbar:2015mca}.  

We have shown that the conditions affecting the development of ELN crossings  are not local in nature. In particular, the appearance of ELN crossings is determined by the slope of the baryon density profile together with the requirement that the $\nu_e$ and $\bar\nu_e$ number densities are comparable in the proximity of the decoupling region. Our simple spherically symmetric SN model hints that  ELN crossings can only occur  in the late stages of the accretion phase and in the cooling phase, under the assumption that a stationary configuration is reached. In fact, at earlier post-bounce times, a large excess of the $\nu_{e}$ number density over the $\bar{\nu}_{e}$ one  prevents ELN crossings from occurring. The latter effect is determined by a baryon density profile that slowly varies with the radius, disfavoring the $\nu_e$ and $\bar\nu_e$ distributions from becoming similar. However, in the late accretion phase and cooling phase the distributions of $\nu_e$ and $\bar\nu_e$ naturally become more similar to each other, $\nu_e$ and $\bar\nu_e$ decouple in closer spatial regions, and favorable conditions for ELN crossings arise.

Due to the numerical challenges involved in solving the equations of motion that include neutrino-neutrino interactions, most of the focus has been on the linear stability analysis of the conditions under which instabilities in the flavor space can occur. However, if flavor instabilities are triggered  in a small localized region of space, is it not clear if and under which conditions the flavor instability would spread, see e.g.~\cite{Capozzi:2017gqd,Yi:2019hrp}. One aspect of the question is whether the flavor evolution changes the neutrino interaction rates therefore leading to a dynamical generation of ELN crossings. 
Our stationary and spherically symmetric SN model suggests that ELN crossings cannot be generated dynamically, unless favorable conditions already exists in the SN core. 

Our model neglects  perturbations coming from global asymmetries induced by hydrodynamic instabilities occurring in SNe. However, it still provides good insights on the generation of ELN crossings under stationary conditions.

A concrete list of necessary and sufficient conditions under which fast pairwise conversions of neutrinos can occur in the SN core   still remains unsettled. Our work provides new insights on the solution of this intriguing jigsaw.

\acknowledgments
We acknowledge insightful discussions with Thomas Janka, Georg Raffelt and Anna Suliga, and are grateful to Robert Bollig for granting access to the data of the $18.6 M_\odot$ SN model adopted in this work.
SS and IT acknowledge support from the Villum Foundation (Project No.~13164). The work of IT has also been supported by the Knud H\o jgaard Foundation and the 
Deutsche Forschungsgemeinschaft through 
Sonderforschungsbereich SFB~1258 ``Neutrinos and Dark Matter
in Astro- and Particle Physics (NDM).

\appendix
\label{appendix}

\begin{figure}
\includegraphics[width=0.49\textwidth]{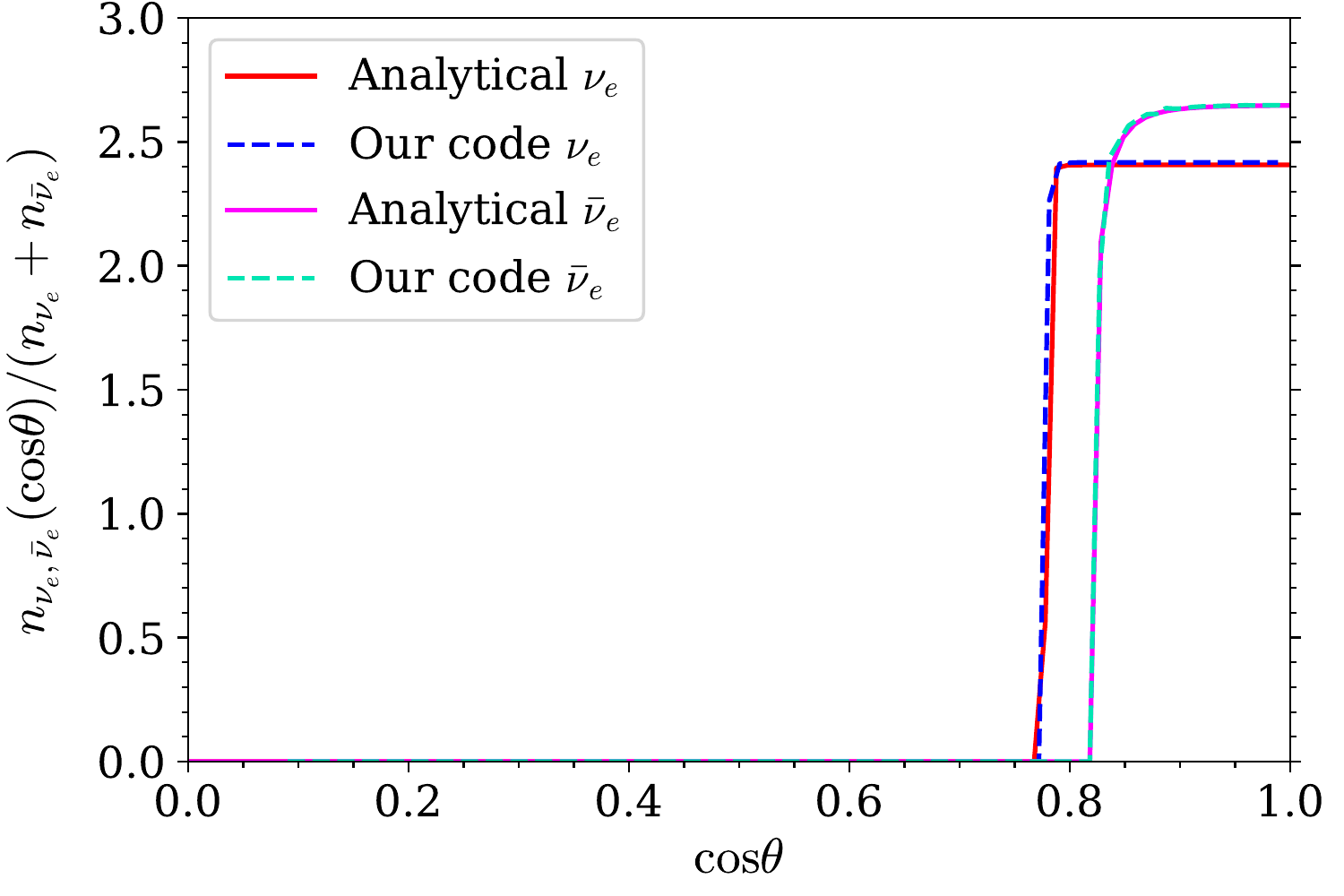}
\includegraphics[width=0.49\textwidth]{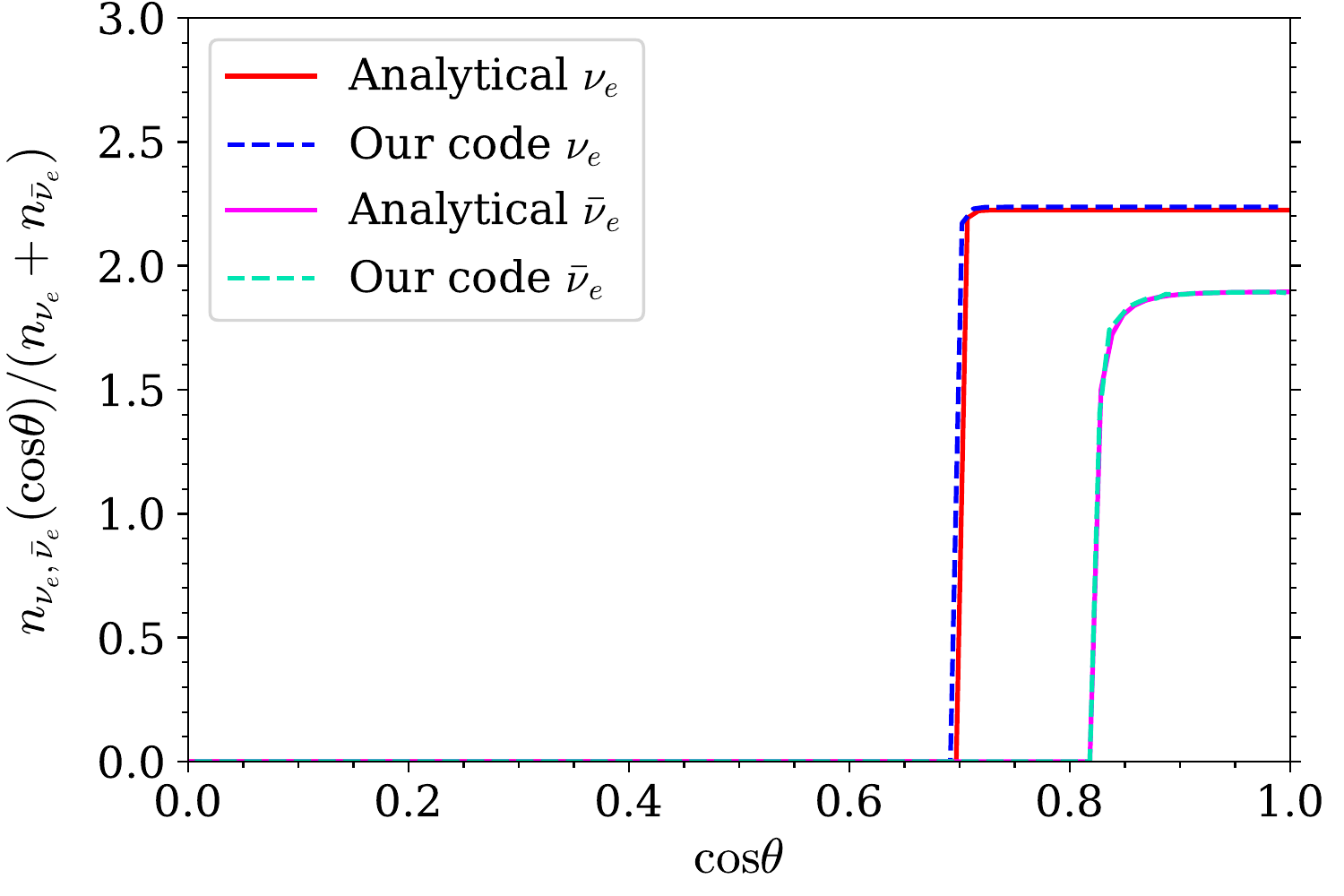}
\caption{Angular distributions of $\nu_{e}$ and $\bar{\nu}_{e}$ for the uniform sphere problem obtained with the analytical approximation derived in~\cite{Murchikova:2017zsy} (continuous lines) and with  our simplified iterative scheme (dashed lines). The angular distributions are plotted at $r=35$~km for $\kappa_{\nu_{e}} = 0.6$~km$^{-1}$ for  and $\kappa_{\bar{\nu}_{e}} = 0.2$ km$^{-1}$. In the top panel, $R_{\nu_{e}} = 22$~km and $R_{\bar{\nu}_{e}} = 20$~km, while in the bottom panel 
$R_{\nu_e} = 25$~km and $R_{\bar{\nu}_e} = 20$~km. The results obtained through the numerical scheme are in good agreement with the analytical approximation.}
\label{analytic}
\end{figure}

In order to further test the validity of our scheme, in this Appendix we adopt the uniform sphere problem as a test case. We consider a sphere emitting uniform black-body radiation and compare the findings of our simplified numerical scheme with the analytical results reported in Sec.~4.2 of~\cite{Murchikova:2017zsy}. 
%We use simple deviations from the black body to provide physical insight in to the problem.
We expect to find a constant intensity across the surface with limb darkening at the edge of the sphere due to the effects of finite optical depth. This is the case of neutrinos emitted from a certain flavor-dependent neutrinosphere. 

To reproduce the uniform sphere case, we assume that electron neutrinos are emitted at a uniform rate inside a sphere of radius $R_{\nu_{e}}$ and electron anti-neutrinos are emitted within a sphere of radius of $R_{\bar{\nu}_{e}}$. As neutrinos and anti-neutrinos propagate outwards they are absorbed at a constant rate, $\kappa_{\nu_e, \bar\nu_e}$, inside $R_{\nu_{e}}$ and $R_{\bar{\nu}_{e}}$ respectively. 
Under these assumptions the total number of neutrinos emitted by the neutrinosphere is 

\begin{eqnarray}
N_{\nu_e, \bar\nu_e} = 4\pi R_{\nu_e,\bar\nu_e}^{2} \int_{0}^{\infty} \frac{E^{2}}{e^{(E-\mu_{\nu_e,\bar\nu_e})/T}+1} \sim R_{\nu_e, \bar\nu_e}^{2}T^{3}\ ;
\end{eqnarray}
while the angular distribution at a distance $r$ from the center of the neutrino-sphere is~\citep{Murchikova:2017zsy}
\begin{eqnarray}
\label{analytic1}
n_{\nu_e, \bar\nu_e}(\cos\theta) \propto \begin{cases}
B \left[1-\exp{\left[g_{\nu_e, \bar\nu_e}(\cos\theta)\right]}\right] \\
\quad \quad \quad \textrm{if} \cos\theta < \sqrt{1-\left(\frac{r}{R_{\nu_e, \bar\nu_e}}\right)^{2}}\\

0\\
\quad \quad \quad \textrm{if} \cos\theta \geq \sqrt{1-\left(\frac{r}{R_{\nu_e, \bar\nu_e}}\right)^{2}}\ ,
\end{cases}
\end{eqnarray}
where $B \simeq R_{\nu_e, \bar\nu_e}^2/T^3$, and
\begin{eqnarray}
g_{\nu_e, \bar\nu_e}(\cos\theta) = -2 \kappa_{\nu_e, \bar\nu_e} R_{\nu_e, \bar\nu_e}  \sqrt{1-\left(\frac{r}{R_{\nu_e, \bar\nu_e}}\right)^{2}(1-\cos^{2}\theta)}\ .
\end{eqnarray}

The continuous lines shown in Fig.~\ref{analytic} represent the angular distributions  obtained  from Eq.~\ref{analytic1} with temperatures of $\nu_{e}$ and $\bar{\nu}_{e}$ set to $10.0$ and $11.0$~MeV, respectively, and $\kappa_{\nu_{e}} = 0.6$~km$^{-1}$, $\kappa_{\bar{\nu}_{e}} = 0.2$ km$^{-1}$. For the top panel, we set $R_{\bar{\nu}_{e}} = 20$~km and $R_{{\nu}_{e}} = 22$~km; while for the case in the bottom panel, we adopt $R_{\bar{\nu}_{e}} = 20$~km and $R_{{\nu}_{e}} = 25$~km.

We now want to reproduce the same result by adopting the numerical iterative scheme introduced in this paper in order to test the robustness of our method.  Note that, while in the rest of the paper we assume that the total number of particles is conserved within the sphere of radius $r_{\mathrm{max}}$, to reproduce the uniform sphere problem we need to take a constant $\mathcal{C}^{\mathrm{gain}}$ term in Eq.~\ref{geneom1} which does not depend on $\rho$ and $\bar{\rho}$ for $r \le R_{{\nu}_{e},\bar{\nu}_{e}}$ and it is zero otherwise. The $\mathcal{C}^{\mathrm{loss}}$ term stays unchanged for  $r \le R_{{\nu}_{e},\bar{\nu}_{e}}$ and it is zero at larger radii. The absorption coefficient introduced above corresponds to $\lambda_{\nu_e, \bar\nu_e} \equiv 2\kappa_{\nu_e, \bar\nu_e}$. 

The dashed lines in Fig.~\ref{analytic} show the correspondent angular distributions obtained by solving Eq.~\ref{geneom1} through our iterative scheme with the same inputs adopted for the analytical model described above. One can see that our numerical scheme reproduces the analytical result with less than $1\%$ error.
Moreover, a comparison between the top and the bottom panel suggests that we obtain an ELN crossing when the two decoupling radii do not differ much from each other;  the ELN crossing disappears when the ratio of the two decoupling radii is significantly larger than unity.

\bibliographystyle{apj}
\bibliography{crossings1}

\end{document}